\documentclass[10pt,prd,aps,showpacs,nofootinbib,superscriptaddress,floatfix,nopreprintnumbers]{revtex4-2}

\usepackage{amsmath,graphicx,mathrsfs,amsfonts,mathtools}
\usepackage{bm} 
\usepackage{subfigure}
\usepackage[colorlinks=true,linktocpage=true,linkcolor=blue,citecolor=blue]{hyperref}
\usepackage[usenames,dvipsnames]{color}

\begin{document}

\title{
Quantum free-streaming: out of equilibrium expansion for the free scalar fields}
\author{L. Tinti}
\email{dr.leonardo.tinti@gmail.com}

\affiliation{Institute of physics, ul. Uniwersytecka 7, 25-406 Kielce, Poland}

\begin{abstract}

The collisionless Boltzmann equation, also called free-streaming, is a convenient approximation. It is rather simple to implement numerically, and and it is effective at reducing the irregularities of rough initial conditions. It can be obtained as a small $\hbar$ limit from a free scalar quantum field. Namely, by neglecting the $\hbar^2$ term in the dynamical evolution of the Wigner distribution, the quantum precursor of the distribution function.

In this work it is presented the general form for the exact solutions of the Wigner distribution, for a scalar field undergoing a $(1+1)$-dimensional expansion. Namely, with the symmetry constraint of rotation and translation invariance in the transverse plane. It is very different from the (on-shell) free-streaming of classical particles. It is shown how to recognize the classical, $\hbar\to 0$, limit from the general form. The numerical analysis for a specific example shows very large quantum corrections to the classical free-streaming.

\end{abstract}



\maketitle

\section{Introduction}

Relativistic hydrodynamics and transport play a fundamental role in the standard description of heavy-ion collisions~\cite{Kurkela:2018wud,Schlichting:2019abc,Florkowski_2018,Reichert_2022,martin2022influence,PhysRevC.79.044915}. In particular the collisionless Boltzmann equation (free-streaming) is used to connect the very initial stages of the collision to the hydrodynamic phase~\cite{PhysRevC.79.044915,PhysRevC.94.024907,Nijs:2020roc,PARKKILA2022137485}. Essentially, providing the initial conditions for the stress-energy tensor $T^{\mu\nu}$, at a finite proper time after the collision. The relativistic Boltzmann equation can be obtained from an effective quantum field theory as a semi-classical expansion~\cite{degroot}. In particular, from a strictly non-interacting scalar field, one obtains the free-streaming equation. The Plank constant, however, is a dimensional quantity. Being $\hbar c\simeq 200$ MeV$\cdot$fm, it is reasonable to expect small quantum corrections in the later stages, especially after the freeze-out. On the other hand at the initial stages, with a temperature of the order of hundreds of MeV, but with the densities changing over a small fraction of a fm, the typical action scale can be of the order of $\hbar$ or even smaller.

It has already been shown that the stress-energy tensor of a quantum gas (free scalar or spinor field) has quantum corrections at global equilibrium\footnote{Just for the non-homogeneous ensembles. Namely, equilibrium with rotation and/or acceleration.There are no corrections for the (grand-)canonical ensemble.}~\cite{Becattini:2011ev,Becattini:2015nva, Becattini:2020qol,Palermo:2021hlf, Ambrus:2019cvr}, and there is no reason to expect no corrections at all out of equilibrium. It bears mentioning, even if it is not the focus of this work, that the local equilibrium versions of such global equilibrium distributions, have been crucial to explain the polarization of particles in peripheral heavy-ion collisions~\cite{Becattini:2013fla, Becattini:2021suc, Becattini:2021iol}. Perhaps understandably, since particles' spin is itself an $\hbar$ correction to the orbital angular momentum, which in turn can be arbitrary large.

More recently, it has been computed exactly the $(0+1)-$dimensional expansion of a scalar field with local equilibrium initial conditions~\cite{Akkelin:2018hpu,Akkelin:2020cfs,Rindori:2021quq}. The treatment is restricted to the so called Bjorken symmetry. Namely, longitudinal boost invariance, homogeneity in the transverse plane (translation and rotation invariance), and axis reflection invariance. These works highlight non-trivial vacuum effects, even for a free scalar theory~\cite{Akkelin:2018hpu,Akkelin:2020cfs}, and significant deviations from the classical free-streaming. Deviations that do not vanish, even in the long time limit~\cite{Rindori:2021quq}.

The purpose of this work is to present a novel class of exact solutions for the non-equilibrium expansion of a scalar field. The symmetry is relaxed to just the homogeneity in the transverse plane, no boost invariance and axis reflection invariance. The requirement of local equilibrium initial conditions is also relaxed.

In Section~(\ref{sec:off}) there is a presentation of the Wigner distribution for a free scalar field, the main subject of this work. It is also explained why, in a dynamical situation, one expects off-shell contributions. In Section~(\ref{sec:1+1}) it is presented the general class of solutions for the $(1+1)$-dimensional expansion. In Section~(\ref{sec:semiclassical}) It is presented the semiclassical limit, and there are the numerical comparisons. The final remarks and conclusions are in Section~(\ref{sec:conclusions}).

In this work, unless otherwise noted, the natural units are in use $\hbar=c=k_{\rm B}=1$ .Repeated indices are assumed to be contracted, and the ``mostly-minus'' convention is adopted for the Minkowski metric 
$g_{\mu \nu}={\rm diag}(1,-1,-1,-1)$. Contractions are indicated with a dot $v^\mu \omega_\mu=v\cdot \omega$. Tree-vectors (in the lab frame) are indicated with bold letters $p=(p^0,{\bf p})$. The D'Alembert operator is represented by a box $\Box = \partial_\mu\partial^\mu$, and $f(x)\overset{\leftrightarrow}{\partial_\mu}g(x) = f(x)\partial_\mu g(x)-[\partial_\mu f(x)]g(x)$ stands for the ``right minus left'' derivative.

\section{The off-shell Wigner distribution}
\label{sec:off}

In Refs.~\cite{Akkelin:2018hpu,Akkelin:2020cfs,Rindori:2021quq} the starting point is the precise knowledge of the state of the system, that is, the density matrix $\hat \rho$. To be more precise, the state chosen therein coincides with the local equilibrium~\cite{JOUR,VANWEERT1982133,Becattini:2014yxa,PhysRevD.92.065008,Harutyunyan:2021rmb} at the time $\tau_0$. Knowing the state of the system one can indeed compute the exact expectation value of the stress-energy tensor 
\begin{equation}
    T^{\mu\nu} = {\rm tr}\left( \hat{\rho} \; \widehat{T}^{\mu\nu} \right).
\end{equation}
It is a straightforward task, if long and tedious for non-equilibrium states. It can be performed exactly, for simple enough configurations, as it has been shown in Refs.~\cite{Akkelin:2020cfs,Rindori:2021quq}.

In this work it is assumed less knowledge of the system. Following Ref.~\cite{degroot}, the Wigner distribution

\begin{equation}\label{Wigner_charged}
    W(x,k)= 2\int \frac{d^4v}{(2\pi)^4} \;  e^{-ik\cdot v} \; {\rm tr}\left( \hat{\rho} : \widehat{\phi}^\dagger(x+\tfrac{v}{2}) \widehat{\phi}(x-\tfrac{v}{2}): \right),
\end{equation}
is enough to obtain the expectation values of the stress-energy tensor

\begin{equation}\label{Tmunu_charged}
    T_{\mu\nu}(x) = {\rm tr}\left( \hat{\rho} \; \widehat{T}_{\mu\nu}(x) \right)= -\frac{1}{2}{\rm tr}\left( \hat{\rho} :\widehat{\phi}^\dagger(x) \overset{\leftrightarrow}{\partial_\mu}\overset{\leftrightarrow}{\partial_\nu}\widehat{\phi}(x):\right) = \int d^4 k \; k_\mu k_\nu \, W(x,k),
\end{equation}
and the conserved current

\begin{equation}\label{current}
    J_{\mu}(x) = {\rm tr}\left( \hat{\rho} \; \widehat{J}_{\mu}(x) \right)=  i\, {\rm tr}\left( \hat{\rho} :\widehat{\phi}^\dagger(x) \overset{\leftrightarrow}{\partial_\mu}\widehat{\phi}(x):\right) = \int d^4 k \; k_\mu \, W(x,k),
\end{equation}
 of a free charged field. The generalization to a free uncharged scalar field is immediate. Namely, just dropping the factor of $2$ in the definition~(\ref{Wigner_charged}), in order to preserve the analogous of Eq.~(\ref{Tmunu_charged})

\begin{equation}\label{Tmunu_gen}
    T^{\mu\nu}(x)  = \int d^4 k \; k^\mu k^\nu \, W(x,k),
\end{equation}
since there is an additional $1/2$ at the operator level in  $\widehat{T}^{\mu\nu}$\footnote{Which, in turn, stems from the over all  $1/2$ factor in the Lagrangian of the uncharged field.}.

As long as one's interest is restricted to just $T^{\mu\nu}$ ( and perhaps the current $J^\mu$), there is no need to know the full state of the system ${\hat \rho}$. The Wigner distribution is sufficient. Needless to say, if one is not only interest, eg, in the energy $\int dV T^{00}={\rm tr}\left({\hat \rho}:{\widehat H}:\right)$, but in the energy fluctuations too
 ${\rm tr}\left({\hat \rho}:{\widehat H}::{\widehat H}:\right)$, additional knowledge is required.

 The equations of motion (it is an over-determined set) for the Wigner distribution stem directly from the Klein-Gordon equation of the free field (see the discussion in Appendix~\ref{app:derivation}) They can be written in the compact form

 \begin{equation}\label{Wigner_evolution_compact}
     \left[ \frac{1}{4}\Box -\left( \vphantom{\frac{}{}} k^2- m^2 \right) +ik\cdot\partial\right]W(x,k) =0,
 \end{equation}
 for the uncharged as well as the charged field. Writing explicitly the $\hbar$ and $c$ constants

 \begin{equation}\label{Wigner_evolution_compact_hbar}
     \left[ \frac{1}{4}\hbar^2\Box -\left( \vphantom{\frac{}{}} k^2- m^2c^2 \right) +i\hbar k\cdot\partial\right]W(x,k) =0.
 \end{equation}
By construction the Wigner distribution~(\ref{Wigner_charged}) is real.  Without loss of generality one can separate the purely real and purely imaginary part of~(\ref{Wigner_evolution_compact_hbar}), obtaining the two equations

\begin{equation}\label{Wigner_evolution}
    \hbar^2\Box \, W(x,k) =4\left( \vphantom{\frac{}{}} k^2- m^2c^2 \right)W(x,k), \qquad 
    k\cdot\partial \, W(x,k) =0.
\end{equation}
Neglecting the second order term in $\hbar$, the first equation becomes just the on-shell condition, and one can approximate $W\propto \delta(k^2=m^2)$. The second equation in~(\ref{Wigner_evolution}) is already identical in the form to the collisionless Boltzmann equation, without any approximation. It is thus reasonable to expect that that, keeping the lowest order terms in an $\hbar$ expansion, one recovers the classical free-streaming from the full solutions of the Wigner distribution.

Besides the classical limits, there is a class of trivial on-shell solutions. Namely, the space-time translation invariant ones, in which $\partial_\mu W =0$ everywhere. The first equation of~(\ref{Wigner_evolution}) implies the on-shell condition for translation-invariant states, regardless of of their action scale or any $\hbar$ expansion. The second one is clearly fulfilled because of the symmetry constraints alone. 

More importantly, most of the solutions of interest are either trivial, or off-shell. For example, if one had

\begin{equation}\label{almost_trivial}
    W(x,k)= \delta(k^2-m^2) W_{\rm on}(x,k)=\delta(k^2-m^2)\int\frac{d^4\xi}{(2\pi)^2}\; e^{ix\cdot\xi} \;{\tilde W }_{\rm on}(\xi, k),
\end{equation}
for a generic, Fourier transformable, on-shell function $W_{\rm on}(x,k)\xleftrightarrow{\rm Fourier}{\tilde W }_{\rm on}(\xi, k)$. Then the exact equations~(\ref{Wigner_evolution}) imply

\begin{equation}\label{on_shell_wigner_evoltution}
   k\cdot \xi \;{\tilde W}_{\rm on}(\xi,k)=0, \qquad \xi^2\; {\tilde W}_{\rm on}(\xi,k) =0.
\end{equation}
That is, both $k^\mu$ and $\xi^\mu$ must be on-shell vectors, $k^2=m^2$ and $\xi^2=0$, and they must be orthogonal. However, non-vanishing light-like and time-like vectors are never orthogonal. A solution proportional to $\delta^4(\xi)$ fulfills both equations, and it is trivial. According to~(\ref{almost_trivial}), the dependence on $x$ is lost. Additional information on the subject is included in Appendix~\ref{app:constraints}.

Since it is so common to have an off-shell Wigner distribution. It is particularly important to have some non-trivial exact solutions see how large can be the quantum corrections to the classical free-streaming.

\section{The (1+1)-dimensional expansion}
\label{sec:1+1}

A general classification of the solutions of Eqs.~(\ref{Wigner_evolution}) is beyond the scope of the present work. Many times, if the full $(3+1)$-dimensional case is beyond the reach of investigation, it is convenient to insert symmetries that reduce the effective dimensions. That is, the number of parameters to take into account in the partial differential equations. In the context of heavy-ion collisions the so called ``Bjorken symmetry'' and its generalization~\cite{Bjorken:1982qr,Gubser:2010ze} are of particular interest. The relativistic Boltzmann equation has been solved exactly for such symmetries~\cite{BANERJEE198916,PhysRevLett.113.202301,Florkowski:2014txa}. These solutions have been used to check how accurate are the approximations schemes used in Heavy-ion collisions~\cite{Florkowski:2013lya,Florkowski:2015cba,Bazow:2015cha,Nopoush:2014qba,Tinti:2015xra,PhysRevD.94.125003,Tinti:2018qfb,Jaiswal:2021uvv}. At least for a solved (and solvable) case, which shares some of the characteristics expected in the real plasma. Like the high energy density and fast longitudinal expansion.

Given the popularity in the literature, it seems reasonable to start from the ``Bjorlken symmetry''. That is, invariance under rotation and translations in the transverse plane (the euclidean group in two dimensions ${\rm ISO}(2)$), under longitudinal boosts (${\rm SO(1,1)}$) and under axis reflection (${\rm \bf{Z}}_2$) in the longitudinal direction (hence full parity invariance, thanks to the rotation invariance in the transverse plane). Surprisingly enough, it is simpler to to find the form of the solutions for a smaller symmetry group. It is always possible implement later the additional symmetries. In this section, it will be considered just  the ${\rm ISO}(2)$ part of the full ``Bjorken symmetry'' ${\rm ISO(2)}\otimes{\rm SO(1,1)}\otimes{\rm \bf{Z}}_2$, the homogeneity in the transverse plane.

In general, the invariance of the density matrix $\hat \rho$ with respect an element of the Poincar\'e group constraints the Wigner distribution in the Following way: Let's call ${\hat {\sf T}}(\Lambda)$ and ${\hat {\sf T}}(a)$ the operator representations of, respectively, a Lorentz transformation $\Lambda$ and a space-time translation $a$. The invariance of the state with respect to either of the  ${\hat {\sf T}}$

\begin{equation}
    {\hat {\sf T}} \; {\hat \rho} \; {\hat {\sf T}}^{-1} = {\hat \rho},
\end{equation}
implies that

\begin{equation}
   {\rm tr}\left( \hat{\rho} : \widehat{\phi}^\dagger(x_2) \widehat{\phi}(x_2): \right) = {\rm tr}\left( {\hat {\sf T}} \; {\hat \rho} \; {\hat {\sf T}}^{-1} : \widehat{\phi}^\dagger(x_1) \widehat{\phi}(x_2): \right).
\end{equation}
Making use of the cyclic property of the trace, and the transformation properties of scalar fields~\cite{Peskin:1995ev}, one has then

\begin{equation}\label{invariance}
    \begin{split}
        {\rm tr}\left( \hat{\rho} : \widehat{\phi}^\dagger(x_1) \widehat{\phi}(x_2): \right) &= {\rm tr}\left( {\hat \rho} \; :{\hat {\sf T}}^{-1} \widehat{\phi}^\dagger(x_1) \widehat{\phi}(x_2) {\hat {\sf T}}: \right)= {\rm tr}\left( {\hat \rho} \; :{\hat {\sf T}}^{-1} \widehat{\phi}^\dagger(x_1) {\hat {\sf T}} \; {\hat {\sf T}}^{-1} \widehat{\phi}(x_2) {\hat {\sf T}}: \right)=\\
       &= {\rm tr}\left( \hat{\rho} : \widehat{\phi}^\dagger(x_1^\prime) \widehat{\phi}(x_2^\prime): \right),
    \end{split}
\end{equation}
the primed coordinates being either $x^\prime = \Lambda^{-1}\cdot x$, or $x^\prime= x-a$, respectively for ${\hat {\sf T}}(\Lambda)$ and ${\hat {\sf T}}(a)$. Plugging this results in the definition~(\ref{Wigner_charged}) one has, first, that the translations do not affect the $v^\mu$ integration variables, just the position $x^\mu$. In particular, if the state is invariant with respect to translations in the transverse plane, the Wigner distribution must fulfill

\begin{equation}
    W(t,x,y,z,k) = W(t,x+\Delta x ,y +\Delta y,z,k), \qquad \forall \Delta x,\Delta y.
\end{equation}
it must depend only on the time and longitudinal direction then $W(t,z,k)$.

In the case Lorentz transformations, on the other hand, the $v^\mu$ do change. The invariance of the state under a Lorentz transformation $\Lambda$ implies\footnote{In the last to last passage has been used the definition of the Lorentz transformations. That is, the transformations that leave the tensor $g_{\mu\nu}$ invariant. For $\Lambda^{-1}$, $(\Lambda^{-1})^\alpha_\mu (\Lambda^{-1})^\beta_\nu g_{\alpha\beta}= g_{\mu\nu}$. It has also been used, in the last passage, the fact that Lorentz transformations have a unitary Jacobian determinan.}

\begin{equation}
    \begin{split}
        &\int d^4v \;  e^{-ik\cdot v} \; {\rm tr}\left( \hat{\rho} : \widehat{\phi}^\dagger(x+\tfrac{v}{2}) \widehat{\phi}(x-\tfrac{v}{2}): \right) = \int d^4v \;  {\exp}\left\{-ik\cdot v\right\} \; {\rm tr}\left( \hat{\rho} : \widehat{\phi}^\dagger(\Lambda^{-1}\cdot x+\tfrac{\Lambda^{-1}\cdot v}{2}) \widehat{\phi}(\Lambda^{-1}\cdot x-\tfrac{\Lambda^{-1}\cdot v}{2}): \right)\\
        & \qquad = \int d^4v \;  {\exp}\left\{-i\left( \Lambda^{-1}\cdot k\right)\cdot \left( \Lambda^{-1}\cdot v \right) \right\} \; {\rm tr}\left( \hat{\rho} : \widehat{\phi}^\dagger(\Lambda^{-1}\cdot x+\tfrac{\Lambda^{-1}\cdot v}{2}) \widehat{\phi}(\Lambda^{-1}\cdot x-\tfrac{\Lambda^{-1}\cdot v}{2}): \right)
        \\
        & \qquad \qquad =\int d^4v \;  {\exp}\left\{-i\left( \Lambda^{-1}\cdot k\right)\cdot v \right\} \; {\rm tr}\left( \hat{\rho} : \widehat{\phi}^\dagger(\Lambda^{-1}\cdot x+\tfrac{ v}{2}) \widehat{\phi}(\Lambda^{-1}\cdot x-\tfrac{v}{2}): \right)
    \end{split}
\end{equation}
and therefore $W(x,k)=W(\Lambda^{-1}\cdot x, \Lambda^{-1}\cdot k)$, as one might expect.

Finally, as a consequence of the invariance of the density matrix $\hat \rho$ under rotations and translations in the transverse plane, the Wigner distribution mus read

\begin{equation}\label{symmetry-simplified_Wigner}
    W(x,k) = W(t,z;k^0, k_T, k^z),
\end{equation}
that is, it depends on the modulus $k_T=\sqrt{(k^x)^2+(k^y)^2}$ but not on the direction in the transverse plane. The second equation in~(\ref{Wigner_evolution}) is relatively straightforward to solve thanks to the symmetry constraints~(\ref{symmetry-simplified_Wigner}). Making use of the method of characteristics~\cite{evans10} one gets $W_f(zk^0-tk^z;k^0,k_T,k^z)$ as the only (continuous, differentiable) function valued solutions. It is immediate to notice that this class of solutions can be generalized to include some distribution-valued solutions. Namely, $\delta(k^0)W_{0}(zk^0-tk^z;k^0,k_T,k^z)=\delta(k^0)W_{0}(-tk^z;k^0,k_T,k^z)$ and $\delta(k^z)W_{T}(zk^0-tk^z;k^0,k_T,k^z)=\delta(k^z)W_{T}(zk^0;k^0,k_T,k^z)$. However among the distribution valued solution, ther is and additional one to consider. The most general solution has the form

\begin{equation}
    W(t,z;k^0,k_T,k_z) = W_r(zk^0-tk^z;k^0,k_T,k^z) + \delta(k^0)\delta(k^z) W_d (t,z;k_T).
\end{equation}
The $t$ and $z$ dependence of $W_d(t,z;k_T)$, at the moment, is unconstrained. It will be specified by the first of the equatios in~(\ref{Wigner_evolution}). The regular part $W_r(zk^0-tk^z;k^0,k_T,k^z)$ can be also proportional to $\delta(k^0)\delta^(k^z)$, in general. This is less interesting, though, because the space-time dependence disappears, but it is acceptable. 

In accordance with the previous literature~\cite{Florkowski:2013lya} it is convenient to introduce the notation

\begin{equation}\label{v_w_def}
    w= zk^0-tk^z, \qquad v = tk^0-z k^z,
\end{equation}
the first one eases the notation in $W_r$, the $v$ will be useful later. 
The first equation in~(\ref{Wigner_evolution}) then reads, for the regular part $W_r$

\begin{equation}
    \left( \vphantom{\frac{}{}} (k^0)^2 -(k^z)^2\right)W_r^{\prime\prime}(w;k_0,k_T,k^z) = -4\left( \vphantom{\frac{}{}} k^2 -m^2 \right)W_r(w;k_0,k_T,k^z),
\end{equation}
the primes in $W_r^{\prime\prime}$ standing for the derivatives with respect to $w$: $W^{\prime\prime}_r = \partial^2_w W_r$.

The general solution of the last equation reads

\begin{equation}
    \begin{split}
        W_r(w;k_0,k_T,k^z) =& \cos\left( 2 w\sqrt{\frac{k^2-m^2}{(k^0)^2 -(k^z)^2}} \right) \; {\cal F}_{\rm even} (k_0,k_T,k^z) \\
        & + \sqrt{\frac{(k^0)^2 -(k^z)^2}{k^2-m^2}} \; \sin\left( 2 w\sqrt{\frac{k^2-m^2}{(k^0)^2 -(k^z)^2}} \right) \; {\cal F}_{\rm odd} (k_0,k_T,k^z).
    \end{split}
\end{equation}
The additional square root in the second term is to ensure a smooth transition of the sine function, when the radical becomes immaginary. Which is a possibility for $k^2<0$. A space-like $k^\mu$,in fact, cannot be excluded in general (see the Appendix~\ref{app:constraints} for additional information). The cosine already goes smoothly from $\cos(x)$ to $\cosh(x)= \cos (ix)$ (hence real) if the argument becomes purely imaginary.

The first equation in~(\ref{Wigner_evolution}) can be solved rather immediately  for the residual $W_d$ since it is, essentially the Klein-Gordon equation of a two-dimensional scalar field with mass equal to $2\sqrt{m^2 +k_T^2}$

\begin{equation}
    \left( \vphantom{\frac{}{}} \partial_t^2 -\partial_z^2 \right)W_d(t,z;k_T) = - 4\left( \vphantom{\frac{}{}} m^2+k_T^2 \right)W_d(t,z;k_T).
\end{equation}
Calling $m_T = \sqrt{m^2 +k_T^2}$ the transverse mass, the general solution of the last equation reads

\begin{equation}
    W_d(t,z;k_T) = \int d\xi \left[  \vphantom{\frac{}{}} e^{-i\left( \vphantom{\frac{}{}} t \sqrt{ 4m_T^2 +\xi^2} - z\, \xi \right) } {\cal A}(\xi;k_T) + e^{i\left( \vphantom{\frac{}{}} t \sqrt{ 4m_T^2 +\xi^2} - z\, \xi \right) } {\cal A}^*(\xi;k_T)\right].
\end{equation}
Finally, summing all up, the Wigner distribution of a state invariant by translations and rotations in the transverse plane reads

\begin{equation}\label{gen_sol_1+1}
    \begin{split}
        &W(t,z;k^0,k_T,k^z) = \delta(k^0)\delta(k^z)\int 
d\xi \left[  \vphantom{\frac{}{}} e^{-i\left( \vphantom{\frac{}{}} t \sqrt{ 4m_T^2 +\xi^2} - z\, \xi \right) } {\cal A}(\xi;k_T) + e^{i\left( \vphantom{\frac{}{}} t \sqrt{ 4m_T^2 +\xi^2} - z\, \xi \right) } {\cal A}^*(\xi;k_T)\right] \\
        & +  \cos\left( 2 w\sqrt{\frac{k^2-m^2}{(k^0)^2 -(k^z)^2}} \right) \; {\cal F}_{\rm even}(k_0,k_T,k^z) + \sqrt{\frac{(k^0)^2 -(k^z)^2}{k^2-m^2}} \; \sin\left( 2 w\sqrt{\frac{k^2-m^2}{(k^0)^2 -(k^z)^2}} \right) \; {\cal F}_{\rm odd}(k_0,k_T,k^z).
    \end{split}
\end{equation}
This kind of solutions can be unexpected. It is very different from the classical free-streaming. The first part in the right hand side is by construction space-like $k^2<0$ and clearly does not correspond to anything in the classical limit. The last two terms can be time-like. However, they do not resemble at all the classical (non-negative) solutions. In fact, in both terms the only space-time dependence is from the $w$ in the trigonometric functions. Regardless of the value, positive or negative, of ${\cal F}_{\rm even}$ and ${\cal F}_{\rm odd}$, by varying $t$ and $z$ one can obtain any value of $w$. Therefore changing the sign of the Wigner distribution, for  any non-vanishing starting point.

On the other hand, the classical solutions are not necessary a bad approximation of the quantum free-streaming in $(1+1)$ dimensions, as it will be shown in next section.

\section{The semiclassical limit}
\label{sec:semiclassical}

Making use of the definitions of $m_T$, $v$ and $w$ (see~(\ref{v_w_def})), and introducing an additional, dimensionless, quantity

\begin{equation}\label{chi}
    \chi = 2 \sqrt{\frac{k^2-m^2}{(k^0)^2-(k^z)^2}}= 2 \sqrt{\frac{k^2-m^2}{k^2+k_T^2}},
\end{equation}
it is possible to rewrite the general solution~(\ref{gen_sol_1+1}) in the compact way

\begin{equation}\label{gen_sol_compact}
    \begin{split}
        W(t,z;k^0,k_T,k^z) &= \delta(k^0)\delta(k^z)\int d\xi \left[  \vphantom{\frac{}{}} e^{-i\left( \vphantom{\frac{}{}} t \sqrt{ 4m_T^2 +\xi^2} - z\, \xi \right) } {\cal A}(\xi;k_T) + e^{i\left( \vphantom{\frac{}{}} t \sqrt{ 4m_T^2 +\xi^2} - z\, \xi \right) } {\cal A}^*(\xi;k_T)\right] \\
        & +  \cos\left(  w \; \chi \right) \; {\cal F}_{\rm even}(k_0,k_T,k^z) + \frac{\sin\left(  w \; \chi \right) }{\chi}\; {\cal F}_{\rm odd}(k_0,k_T,k^z).
    \end{split}
\end{equation}
Plugging back the $\hbar$ and $c$ constants one has

\begin{equation}\label{gen_sol_compact_hbaar}
    \begin{split}
        W &= \delta(k^0)\delta(k^z)\int d\xi \left[  \vphantom{\frac{}{}} e^{-i\left( \vphantom{\frac{}{}} ct \sqrt{ 4m_T^2 +\xi^2} - z\, \xi \right)/\hbar } {\cal A}(\xi;k_T) + e^{i\left( \vphantom{\frac{}{}} ct \sqrt{ 4m_T^2 +\xi^2} - z\, \xi \right)/\hbar } {\cal A}^*(\xi;k_T)\right] \\
        & +  \cos\left(  \frac{w}{\hbar} \; \chi \right) \; {\cal F}_{\rm even}(k_0,k_T,k^z) + \sin\left(  \frac{w}{\hbar} \; \chi \right) \; \frac{{\cal F}_{\rm odd}(k_0,k_T,k^z)}{\chi} .
    \end{split}
\end{equation}
It is tempting to use (improperly) the  Riemann-Lebesgue lemma~\cite{Gradshteyn:1943cpj,zhou2017notes}, and say that the fast oscillating integral in the first term in going to vanish in the small $\hbar$ limit. If this were correct one could also dismiss the last two terms. They are not integrals and they do not approach anything in the $\hbar\to 0$ limit. However, the stress-energy tensor~(\ref{Tmunu_gen}) and the current~(\ref{current}) are integrals. Applying once again (still improperly) the lemma, one would erroneously conclude that both $T^{\mu\nu}$ and $J^\mu$. are vanishing in the classical limit.

All of this does not take properly into account the explicit $\hbar$ factors in the definition of the Wigner distribution~(\ref{Wigner_charged}), which are present for dimensional reasons, as well as the physical dimensions of the fields $\widehat \phi$\footnote{The definition of the fields themselves includes $\hbar$ factors, regardless of the normalization of the ladder operators. This is a consequence of of the canonical quantization $[\widehat{\phi}(x),\widehat{\Pi}(y)] = i\hbar\delta^3({\bf x}-{\bf y})$ at equal times. For the sake of comparisons with the literature, it can be useful to point out that some authors do not use this form of the canonical quantization~\cite{degroot}. In this case some additional $\hbar$ factors appear in the definitions of physical quantities (to maintain their physical dimensions), the fields still have explicit $\hbar$ factors in their definition, leaving unchanged the dimensional considerations.}.

A more sensible choice is to consider~(\ref{Tmunu_gen}),~(\ref{current}) and their classical counterpart

\begin{equation}
    \begin{split}
        T^{\mu\nu}(x) &= \int d^4 k \; k^\mu k^\nu W(x,k)\quad \xrightarrow{\mbox{ small } \hbar \mbox{ }} \quad \int \frac{d^3 p}{(2\pi\hbar)^3 E_{\bf p}/c} \; p^\mu p^\nu \; \left[  f(x,{\bf p}) + \bar{f}(x,{\bf p}) \vphantom{\frac{1}{2}} \right], \\
        J^{\mu} &= \int d^4 k \; k^\mu W (x,k)\quad \xrightarrow{\mbox{ small } \hbar \mbox{ }} \quad \int \frac{d^3 p}{(2\pi\hbar)^3 E_{\bf p}/c} \; p^\mu \; \left[  f(x,{\bf p}) - \bar{f}(x,{\bf p}) \vphantom{\frac{1}{2}} \right].
    \end{split}
\end{equation}
The Plank constants in the right hand side are essential. They must remain even in the non-relativistic limit. The last limits hold if the quantum Wigner distribution fulfills

\begin{equation}\label{proper_small_hbar}
    \lim_{\hbar\to 0}\left[(2 \pi \hbar)^3 \; W(x,k) \vphantom{\frac{1}{2}}\right] \propto \delta(k^2-m^2).
\end{equation}
For the remainder of this work it will be considered only $W^+$

\begin{equation}
    W^+(x,k) = \theta(k^0)\theta(k^2)W(x,k)=\theta(k^0)\theta(k^2-m^2)W(x,k),
\end{equation}
that is, the particle contribution to the full Wigner distribution\footnote{the general decomposition of the Wigner distribution of massive free fields is known~\cite{Palermo:2021hlf,Becattini:2020qol, Becattini:2020sww,Tinti:2020gyh}, including which part takes contribution exclusively from particles, antiparticles, or from both.See also the discussion in Appendix~\ref{app:constraints} for an introduction and a proof of the statements.}. Only the time-like part of $W$ has any chance to be on shell in the limit~(\ref{proper_small_hbar}), and all the arguments can be trivially extended to the antiparticle contribution $W^-$ (defined with a $\theta(-k^0)$ instead of $\theta(k^0)$). The space-like part $\theta(-k^2)W$ either vanishes in the small $\hbar$ limit~(\ref{proper_small_hbar}) or provides a non-trivial quantum term. More in general, neither the $W^+$ and $W^-$ are granted to have exclusively an on-shell limit. State like these would correspond to systems dominated by diffraction-interference, not unreasonable with a limited number of particles and highly entangled states. The rest of this section, however, is dedicated to find a class of solutions that indeed have recognizable on-shell limit. The main goal being to estimate the quantum corrections between the full quantum limit and the, recognized, classical on-shell limit. The first step is to rewrite the particle contribution of~(\ref{gen_sol_compact_hbaar}), without loss of generality, in the following way

\begin{equation}\label{W+_prefactor}
    (2\pi\hbar)^3 \, W^+ = \theta(k^0)\theta(k^2-m^2c^2)\frac{(4-\chi^2)^2}{4m_T^2 \chi}\left[ \cos\left(\frac{w\chi}{\hbar}\right) {\tilde f}_{\rm even}(k^0,k_T,k^z) + \sin\left(\frac{w\chi}{\hbar}\right) {\tilde f}_{\rm odd}(k^0,k_T,k^z)  \right]\frac{A}{2\pi\hbar}.
\end{equation}
The prefactor $(4-\chi^2)^2/(4m_T^2 \chi)$ seems an unnecessary complication, but its usefulness to obtain the correct~(\ref{proper_small_hbar}) limit will become apparent later. The factor $A/(2\pi \hbar)$ is more familiar. It looks like being simply extracted from the definition~(\ref{Wigner_charged}), with an additional action scale on the numerator $A$, which ensures that ${\tilde f}_{\rm even}$ and ${\tilde f}_{\rm odd}$ have no physical dimensions.

In order to show how to extract the free-streaming limit, it is convenient to make some assumptions now. Up to this point, the time and space positions $t$ and $z$ have been completely arbitrary. For practical purposes it is convenient to go on the forward light cone. Namely,  $t>|z|$. This is the region of interest of heavy-ion collisions, the collision itself is usually assumed to take place at the point $t=z=0$ (for instance~\cite{PhysRevC.79.044915,PhysRevC.94.024907,Nijs:2020roc,PARKKILA2022137485,Akkelin:2018hpu,Akkelin:2020cfs,Rindori:2021quq,Bjorken:1982qr,Gubser:2010ze,BANERJEE198916,PhysRevLett.113.202301,Florkowski:2014txa,Florkowski:2013lya,Florkowski:2015cba,Bazow:2015cha,Nopoush:2014qba,Tinti:2015xra,PhysRevD.94.125003,Tinti:2018qfb,Jaiswal:2021uvv}). In this section of space-time $\theta(k^0)\theta(k^2-m^2c^2)$ is equivalent to $\theta(v)\theta(k^2-m^2c^2)$. More importantly instead of using $k^0$ and $k^z$ one can use $v$ and $w$ in the integration measure

\begin{equation}
    \int d^4 k = \int \frac{d^2k_T \; dv \; dw}{\tau^2}, \qquad \forall t>|z|
\end{equation}
with $\tau=\sqrt{c^2t^2 -z^2}$ the space-time proper time\footnote{In natural units the $c$ drops and there is no difference between a lenght and a time interval, at the level of physical dimensions.}. It is possible to make sense of the prefactor in~(\ref{W+_prefactor}) at this point. Making use of the definition of $\chi$~(\ref{chi})

\begin{equation}\label{v_chi}
    \chi^2 = 4 \frac{k^2-m^2c^2}{k^2+k_T^2} = 4\frac{v^2-w^2 + m_T^2\tau^2}{v^2-w^2}, \qquad \Rightarrow\qquad v^2=w^2+\frac{4}{4-\chi^2}m_T^2 \tau^2,
\end{equation}
one can make the following change of variables
\begin{equation}
    \begin{split}
        \int d^4 k \; \theta(k^0) \theta(k^2-m^2c^2) &= \int \frac{d^2k_T \; dv \; dw}{\tau^2}\theta(v)\theta\left( \tfrac{v^2-w^2}{\tau^2} -m_T^2 \right) = \int \frac{d^2k_T \; dw}{\tau^2}\int_0^\infty dv\; \theta\left( \tfrac{v^2-w^2}{\tau^2} -m_T^2 \right)\\
        &= \int \frac{d^2k_T \; dw}{\tau^2}\int_0^2 d\chi \frac{4m_T^2 \tau^2 \chi}{v(4-\chi^2)^2} =\int d^2k_T \; dw \int_0^2 d\chi \frac{4m_T^2}{v(4-\chi^2)^2}\chi,
    \end{split}
\end{equation}
the $v$ being, of course, the positive root of $v^2$ in~(\ref{v_chi}). The awkward prefactor in~(\ref{W+_prefactor}) is there to partially compensate with the Jacobian of the transformation. Therefore, in the forward light cone

\begin{equation}\label{mom_int_pref}
    \begin{split}
        \int d^4 k \; \theta(k^0) \theta(k^2-m^2c^2)\frac{(4-\chi^2)^2}{4m_T^2 \chi} &= \int_0^2 d\chi \int \frac{d^2k_T \; dw}{v} = \int_0^2 d\chi \int \frac{d^2k_T \; dw}{\sqrt{\tfrac{4}{4-\chi^2}m_T^2 \tau^2 +w^2}}.
    \end{split}
\end{equation}
Confronting the right hand side with the results of relativistic kinetic theory in Milne coordinates (for instance in Ref.~\cite{Florkowski:2013lya}) one recognizes the classical (on-shell) Lorentz covariant measure in the momentum space for $\chi=0$. In other words, if the right hand side of~(\ref{W+_prefactor}) becomes proportional to $\delta(\chi)$ in the small $\hbar$ limit, one recovers the classical on-shell integrals. Another way to show that, is to compute exactly

\begin{equation}
    \begin{split}
       &\theta(k^0)\theta(k^2-m^2c^2) \frac{(4-\chi^2)^2}{4m_T^2 \chi}\; \delta(\chi) = \frac{(4-\chi^2)^2}{4m_T^2 \chi}\; \frac{\delta(k^0- \sqrt{m_T^2+(k^z)^2})}{\left| \tfrac{\partial\chi}{\partial k^0} \right|} = \frac{4m_T^2}{(k^2 +k_T^2)^2\chi}\; \frac{\delta(k^0- \sqrt{m_T^2+(k^z)^2})}{\left| \tfrac{\partial\chi}{\partial k^0} \right|} \\
       &\qquad=\frac{4m_T^2}{(k^2 +k_T^2)^2\chi}\; \delta\left(k^0- \sqrt{ m_T^2+(k^z)^2}\right)\;  \frac{\chi (k^2+k_T^2)^2}{4 \, k^0 \, m_T^2} = \frac{\delta(k^0- \sqrt{m_T^2+(k^z)^2})}{k^0}.
    \end{split}
\end{equation}

The main point now is to show which class (or, better, classes) of ${\tilde f}$'s provide such a $\delta(\chi)$ in the proper classical limit~(\ref{proper_small_hbar}). The starting step is to generalize the familiar concept of an $\varepsilon$ dependent $\delta$ family

\begin{equation}\label{deltaepsilon}
    \delta_\varepsilon (x) = \frac{1}{\varepsilon}\; g\left(\frac{x}{\varepsilon}\right) \; \xrightarrow{\;\varepsilon\to 0\;} \; \delta(x),
\end{equation}
assuming the uniform convergence of the integrals of $\delta_\varepsilon(x)$ with the test functions~\cite{wheeden1977measure}. This concept can be extended to the case of a function with a double dependence on $x$, and other parameters $p_1,\cdots$ 

\begin{equation}\label{distribution_limit}
    \frac{1}{\varepsilon}\; g\left(\frac{x}{\varepsilon},x,p_1\cdots\right) \xrightarrow{\varepsilon\to 0} \delta(x) \int dy \; g(y,0,p_1,\cdots).
\end{equation}
The last relation can be proved in the same way that~(\ref{deltaepsilon}) is usually proved. Calling $\psi(x)$ a generic test function, 

\begin{equation}
    \int\frac{dx}{\varepsilon}\; g\left(\frac{x}{\varepsilon},x,p_1\cdots\right) \psi(x) = \int dy \; g\left(y,y\varepsilon,p_1\cdots\right) \psi(y\varepsilon) \xrightarrow{\varepsilon\to 0} \psi(0) \int dy \; g(y,0,p_1,\cdots).
\end{equation}
Finally, if the ${\tilde f}$ are of the form

\begin{equation}\label{ansatz}
     {\tilde f}_{\rm even}= {\tilde f}_{\rm even}\left( \chi \tfrac{A}{\hbar}; k^0,k_T,k^z \right), \qquad {\tilde f}_{\rm odd}= {\tilde f}_{\rm odd}\left( \chi \tfrac{A}{\hbar}; k^0,k_T,k^z \right),
\end{equation}
calling $\varepsilon=\hbar/A$, ${\tilde w}= w/A$, and counting the domain of integration in $\chi$ of the momentum integrals~(\ref{mom_int_pref}), one has\footnote{Reminding that $\theta(y\varepsilon) =\theta(y)$ for positive $\varepsilon$. }

\begin{equation}\label{tildef_limit}
    \begin{split}
        \frac{\theta(\chi)\theta(2-\chi)}{(2\pi \varepsilon)} \cos\left( {\tilde w}\tfrac{\chi}{\varepsilon} \right) \; {\tilde f}_{\rm even}\left( \tfrac{\chi}{\varepsilon}; k^0,k_T,k^z \right)&\xrightarrow{\varepsilon\to 0^+} \frac{1}{2}\delta(\chi)\; \int \frac{d\chi^\prime}{(2\pi)}\cos\left( {\tilde w} \chi^\prime \right) {\tilde f}_{\rm even}\left( \chi^\prime; \sqrt{m_T^2+(k^z)^2},k_T,k^z \right), \\
        \frac{\theta(\chi)\theta(2-\chi)}{(2\pi \varepsilon)} \sin\left( {\tilde w}\tfrac{\chi}{\varepsilon} \right) \; {\tilde f}_{\rm odd}\left( \tfrac{\chi}{\varepsilon}; k^0,k_T,k^z \right)&\xrightarrow{\varepsilon\to 0^+} \frac{1}{2}\delta(\chi)\; \int \frac{d\chi^\prime}{(2\pi)}\sin\left( {\tilde w} \chi^\prime \right) {\tilde f}_{\rm odd}\left( \chi^\prime ; \sqrt{m_T^2+(k^z)^2},k_T,k^z \right),
    \end{split}
\end{equation}
the extension of the $\tilde f$ to the negative $\chi$ region being clearly the even one in the $\chi\leftrightarrow-\chi$ exchange for the ${\tilde f}_{\rm even}$, and the odd one for the ${\tilde f}_{\rm odd}$. The right hand side is essentially a Dirac delta and a counter-Fourier transform. It is possible now to select a (non-unique) class of functions that give the classical free-streaming limit. Namely, ${\tilde f}_{\rm even}$ the real part of a Fourier transform in ${\tilde w}=w/A$ (which depends only on the even part in $w\leftrightarrow-w$), and the ${\tilde f}_{\rm odd}$ the imaginary part (depending only in the odd $w\leftrightarrow-w$ part), of the classical distribution function $f$

\begin{equation}\label{simplest_class}
    \begin{split}
        {\tilde f}_{\rm even}\left( \tfrac{\chi}{\varepsilon}; k_T,k^z \right) = 2{\rm Re}\left[\int \frac{d{\tilde w}^\prime}{(2\pi)}\; f\left( {\tilde w}^\prime;k_T,k^z \right) \; e^{ -i{\tilde w}^\prime \tfrac{\chi}{\varepsilon} }\right], \\
        {\tilde f}_{\rm odd}\left( \tfrac{\chi}{\varepsilon}; k_T,k^z \right) = 2{\rm Im}\left[\int \frac{d{\tilde w}^\prime}{(2\pi)}\; f\left( {\tilde w}^\prime;k_T,k^z \right) \; e^{ -i{\tilde w}^\prime \tfrac{\chi}{\varepsilon} }\right].
    \end{split}
\end{equation}
With this class of functions in~(\ref{W+_prefactor}), the momentum integrals of $W^+$ then read

\begin{equation}\label{quantum_integrals}
    \begin{split}
        &\int d^4k W^+ \left[ \vphantom{\frac{}{}} \cdots \right] = \int_0^2 \frac{d\chi}{(2\pi \hbar)^3} \int \frac{d^2k_T \; dw}{\sqrt{\tfrac{4}{4-\chi^2}m_T^2 \tau^2 +w^2}} \left[ \cos\left(\frac{w\chi}{\hbar}\right) {\tilde f}_{\rm even} + \sin\left(\frac{w\chi}{\hbar}\right) {\tilde f}_{\rm odd}  \right]\frac{A}{2\pi\hbar} \left[ \vphantom{\frac{}{}} \cdots \right] \\
        &=\int_0^2 \frac{d\chi}{(2\pi \hbar)^3} \int \frac{d^2k_T \; d{ w}}{\sqrt{\tfrac{4}{4-\chi^2}m_T^2 \tau^2 +{ w}^2}} \left[ \cos\left({\tilde w}\frac{\chi}{\varepsilon}\right) {\tilde f}_{\rm even} + \sin\left({\tilde w}\frac{\chi}{\varepsilon}\right) {\tilde f}_{\rm odd}  \right]\frac{1}{2\pi\varepsilon} \left[ \vphantom{\frac{}{}} \cdots \right],
    \end{split}
\end{equation}
the dots standing either for $k^\mu$ or $k^\mu k^\nu$, respectively, for the current or the stress-energy tensor.

In the small $\varepsilon$ limit, that is,  large action scale of the system in $\hbar$ units, one recover by construction the classical free-streaming for $f$

\begin{equation}\label{classical_integrals}
    \begin{split}
        &\lim_{\varepsilon->0}\int d^4k W^+ \left[ \vphantom{\frac{}{}} \cdots \right] = \frac{1}{(2\pi \hbar)^3} \int \frac{d^2k_T \; d{\tilde w}}{\sqrt{m_T^2 \tau^2 +{\tilde w}^2}} f({\tilde w};k_T,k^z) \left[ \vphantom{\frac{}{}} \cdots \right] \\
        & \qquad \quad= \frac{1}{(2\pi \hbar)^3} \int \frac{d^2k_T \; dw}{\sqrt{m_T^2 \tau^2 +w^2}} f(\tfrac{w}{A};k_T,k^z)\left[ \vphantom{\frac{}{}} \cdots \right].
    \end{split}
\end{equation}
This class of solutions determined by~(\ref{simplest_class}) can be immediately extended in two ways. The first one starts noting that there is not necessarily only a single action scale $A$, per state of the system (classical as well as quantum). Being the $A$ in~(\ref{W+_prefactor}) the \textit{smallest}, non-vanishing one, and $A_i=A_1,A_2\cdots$ the other (larger) action scales. Calling $r_i$ the ratios $r_=A_i/A$, one can immediately extend~(\ref{ansatz}) to

\begin{equation}\label{ansatz_ri}
     {\tilde f}_{\rm even}= {\tilde f}_{\rm even}\left( \chi \tfrac{A}{\hbar}; k^0,k_T,k^z; r_1,r_2, \cdots \right), \qquad {\tilde f}_{\rm odd}= {\tilde f}_{\rm odd}\left( \chi \tfrac{A}{\hbar}; k^0,k_T,k^z ; r_1,r_2, \cdots \right).
\end{equation}
If one of the larger action scale were the one associated with $\chi$, as in $\chi A_i/\hbar$, one can always use $\chi A_i/\hbar=r_i\chi A/\hbar$ and the last form~(\ref{ansatz_ri}) remains general. The limits~(\ref{tildef_limit}) and the definition~(\ref{simplest_class}) immediately generalize to the classical free-streaming function

\begin{equation}
    f(\tfrac{w}{A};k_T,k^z;r_1,r_2,\cdots),
\end{equation}
in the $\varepsilon\to0$ limit, keeping fixed the ratios $r_i$.

The second extension stems from the fact that the $\delta(\chi)$ in the right hand side of~(\ref{tildef_limit}) replaces any $\chi$ with the number $0$. More precisely, reminding the general~(\ref{distribution_limit}), for any continuous dependence of ${\tilde f}(\frac{\chi}{\varepsilon}; \chi,\cdots)$ on the $\chi$ without the division by $\varepsilon$, one replaces $\chi\to 0$. I particular, one can multiply any of the functions ${\tilde f}$, either the~(\ref{ansatz}) or their generalized version~(\ref{ansatz_ri}), by a function $h(\chi)$, as long as it does not spoil the integration properties and $h(0)=1$.

It is possible then to recognize a very large class of exact solutions among the~(\ref{gen_sol_compact_hbaar}), that all go to the same classical  free-streaming solution $f$ in the small $\hbar$ limit~(\ref{proper_small_hbar}). There is an ample margin for discrepancies between the full quantum solutions~(\ref{gen_sol_compact_hbaar}) and their classical approximations given by the distribution function $f$. Unfortunately, the direct computation of integrals like~(\ref{quantum_integrals}) is much more complicated than the classical version~(\ref{classical_integrals}). Mostly because of the fast oscillations introduced by the trigonometric functions. A general prescription to estimate their difference for some finite $\varepsilon$ is currently lacking, and one has to make a case-by-case analysis. A the end of this section it's presented possibly the simplest example relevant for heavy-ion collisions.

First of all, it's considered the massless case, as a significant simplification. There is no difference then between the transverse mass $m_T$ and $k_T$. Being the Bjorken symmetry~\cite{Bjorken:1982qr} so common and of practical importance, it's considered the limit of longitudinal boost invariance, and axis-reflection invariance. The boost invariance imply that only the boost invariant combination $(k^0)^2-(k^z)^2 = (v^2-w^2)/\tau^2$ can appear in the $\tilde f$. The axis invariance requires that only even functions of $w$ are admitted, since the $z\leftrightarrow -z$ transformation sends $w\leftrightarrow -w$. The same symmetry requirements hold for the classical distribution function $f(\tfrac{w}{A},k_T)$, see also Ref.~\cite{Florkowski:2013lya}. The, asymptotic, free-streaming distribution function chosen for the numerical comparisons is

\begin{equation}\label{asympt_free}
    f(w, k_T)=\frac{\pi^4}{30} \; \exp\left\{-\frac{k_T^2}{2T_0^2} \; -\frac{w^2}{2T_0^2\tau_0^2}\right\}.
\end{equation}
Clearly the action scale is $A=T_0\tau_0$. The numerical prefactor is to ensure that at the proper time $\tau=\tau_0$, the energy density (and pressure and all the $T^{\mu\nu}$) is the one of a gas of bosons with temperature $T_0$. It is not an equilibrium state, but, at least in the classical limit, $T_0$ has the role of an ``initial temperature''. In the sense that it gives the actual energy density of the system at $\tau_0$. The choice of the Gaussian form (instead of, eg, the local equilibrium one) is just because Gaussians are easy to Fourier transform.

Instead of using directly the canonical~(\ref{simplest_class}) for the quantum precursor, it's used the modified form

\begin{equation}\label{quantum_num}
    {\tilde f}_{\rm even} =2\sqrt{2\pi}\; \frac{4}{4-\chi^2} \; \frac{\pi^4}{30} \; \exp\left\{-\frac{k_T^2}{2T_0^2} \frac{4}{4-\chi^2} \; -\frac{\chi^2}{2\varepsilon^2}\right\}.
\end{equation}
It still has the same limit as the canonical one, but it allows to significantly simplify the momentum integrals~(\ref{quantum_integrals})

\begin{equation}\label{quantum_bjorken_0}
    \begin{split}
        &\int d^4k W^+ \left[ \vphantom{\frac{}{}} \cdots \right] = \frac{1}{(2\pi \hbar)^3}\frac{\pi^4}{30}\int_{-2}^2 \frac{d\chi}{\varepsilon\sqrt{2\pi}} \int \frac{d^2k_T \; dw}{\sqrt{\tfrac{4}{4-\chi^2}k_T^2 \tau^2 +w^2}}\frac{4}{4-\chi^2}  \cos\left(\frac{ {\tilde w}\chi}{\varepsilon}\right) \exp\left\{-\frac{k_T^2}{2T_0^2} \frac{4}{4-\chi^2} \; -\frac{\chi^2}{2\varepsilon^2}\right\} \left[\vphantom{\frac{}{}} \cdots\right] \\
        &\qquad =\frac{1}{(2\pi \hbar)^3}\frac{\pi^4}{30} \int_{-2}^2 \frac{d\chi}{\varepsilon\sqrt{2\pi}} \int \frac{d^2p_T \; dw}{\sqrt{p_T^2 \tau^2 +w^2}}  \cos\left(\frac{{\tilde w}\chi}{\varepsilon}\right) \exp\left\{-\frac{p_T^2}{2T_0^2}  \; -\frac{\chi^2}{2\varepsilon^2}\right\} \left[ \vphantom{\frac{}{}} \cdots\right],
    \end{split}
\end{equation}
with the change of variables $p_T^2=4k_T^2/(4-\chi^2)$. Rewriting in terms of the dimensionless variables ${\tilde w}= w/A=w/(T_0\tau_0)$ and $y^2 = p_T^2/T_0^2$, and performing the angular integral in the transverse plane

\begin{equation}\label{quantum_bjorken_1}
    \begin{split}
        &\int d^4k W^+ \left[ \vphantom{\frac{}{}} \cdots \right] = \\
        &=\frac{T_0^2}{(2\pi \hbar)^3}\frac{\pi^5}{15} \; \frac{\tau_0}{\tau}{}\int_{-2}^2 \frac{d\chi}{\varepsilon\sqrt{2\pi}} \int_{-\infty}^\infty d{\tilde w}\int_0^\infty \frac{dy}{\sqrt{y^2 +{\tilde w}^2\tfrac{\tau_0^2}{\tau^2}}} \; y\;  \cos\left(\frac{{\tilde w}\chi}{\varepsilon}\right) \exp\left\{-\frac{y^2}{2}  \; -\frac{\chi^2}{2\varepsilon^2}\right\} \left[ \vphantom{\frac{}{}} \cdots\right]
    \end{split}
\end{equation}
The longitudinal pressure (see, for instance, Refs.~\cite{Bjorken:1982qr,Florkowski:2013lya,Rindori:2021quq,Tinti:2020pyb})

\begin{equation}
    {\cal P}_L = z_\mu z_\nu T^{\mu\nu} = \int d^4 k \frac{w^2}{\tau^2} W, \qquad z^\mu =(z,0,0,t)/\tau
\end{equation}
then reads

\begin{equation}\label{PL_int}
    \begin{split}
        {\cal P}_L &=\int d^4k W^+ \left[ \vphantom{\frac{}{}} \cdots \right] = \\
        &=\frac{T_0^4}{(2\pi \hbar)^3}\frac{\pi^5}{15} \; \left(\frac{\tau_0}{\tau}\right)^3\int_{-2}^2 \frac{d\chi}{\varepsilon\sqrt{2\pi}} \int_{-\infty}^\infty d{\tilde w}\int_0^\infty \frac{dy}{\sqrt{y^2 +{\tilde w}^2\tfrac{\tau_0^2}{\tau^2}}} \; y\;  \cos\left(\frac{{\tilde w}\chi}{\varepsilon}\right) \exp\left\{-\frac{y^2}{2}  \; -\frac{\chi^2}{2\varepsilon^2}\right\} \; {\tilde w}^2
    \end{split}
\end{equation}
%
\begin{figure}
	\includegraphics[scale=1.15]{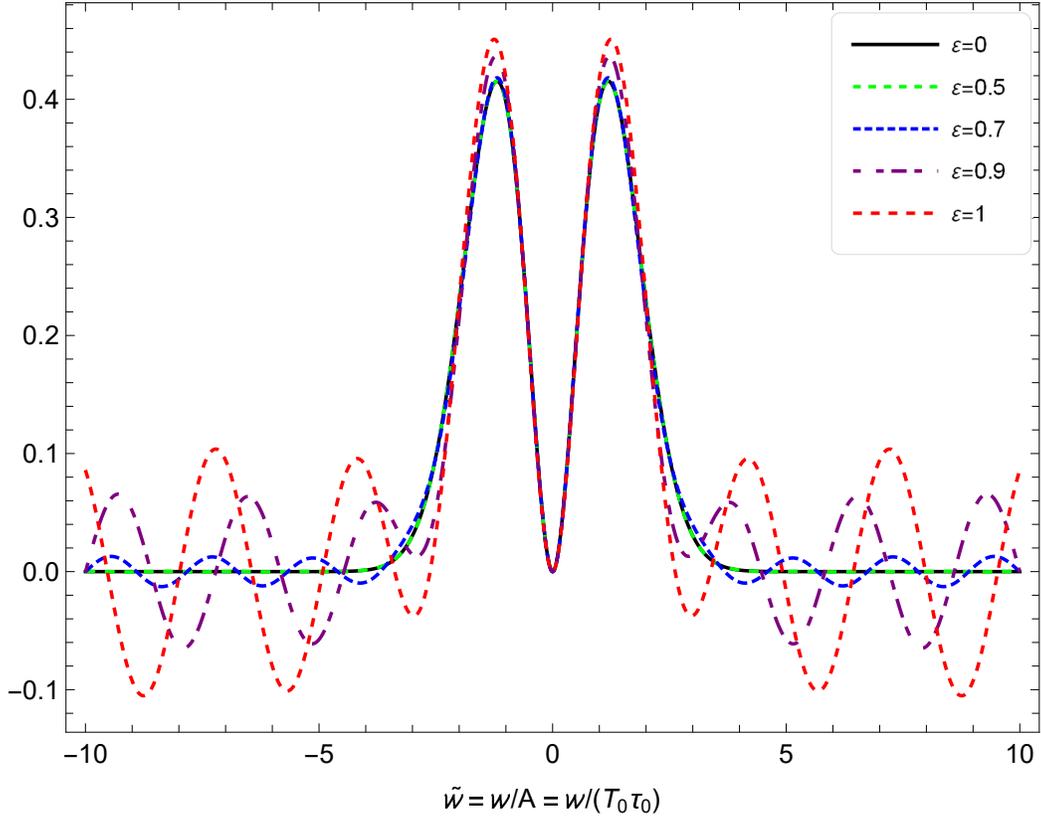}
	\caption{(Color online) Comparison between the integrands of Eq.~(\ref{quantum}) and Eq.~(\ref{classical}), the proper time is $\tau=\tau_0$. The values of
	$\varepsilon=T_0\tau_0/\hbar$ are, respectively, $\varepsilon=0.5$ for the green dotted line, $\varepsilon=0.7$ for the blue dashed line, $\varepsilon=0.9$ for the purple dot-dashed line and $\varepsilon=1$ for the red dashed one. The black solid line being the classical ($\varepsilon\to 0$) result. } 
	\label{Fig_comp}
\end{figure}
The $\chi$ integration, as well as the $y$ integration, can be performed exactly. Removing the inessential $\frac{T_0^4}{(2\pi \hbar)^3}\frac{\pi^5}{15} \; \left(\frac{\tau_0}{\tau}\right)^3$ (which appears, exactly the same, also in the classical integral) one has

\begin{equation}\label{quantum}
    \int_{-\infty}^\infty d{\tilde w} \; \sqrt{\frac{\pi}{2}}\;  \exp\left\{\frac{{\tilde w}^2\tau_0^2}{2\tau^2}\right\} \; {\rm Erfc}\left( \frac{{\tilde w}^2\tau_0^2}{2\tau^2} \right) \; \frac{1}{2}\; \exp\left\{\frac{-{\tilde w}^2}{2}\right\} \left[ {\rm Erf}\left( \frac{2-i\varepsilon {\tilde w}}{\varepsilon \sqrt{2}}\right)+{\rm Erf}\left( \frac{2+i\varepsilon {\tilde w}}{\varepsilon \sqrt{2}}\right) \right].
\end{equation}
This integral is extremely complicated compared to the classical counterpart (its $\varepsilon\to 0$ limit)

\begin{equation}\label{classical}
    \int_{-\infty}^\infty d{\tilde w} \; \sqrt{\frac{\pi}{2}}\;  \exp\left\{\frac{{\tilde w}^2\tau_0^2}{2\tau^2}\right\} \; {\rm Erfc}\left( \frac{{\tilde w}^2\tau_0^2}{2\tau^2} \right) \; \ \exp\left\{-\frac{{\tilde w}^2}{2}\right\} = \frac{2\tau_0\tau}{\tau^2-\tau_0^2} +2\frac{\tau_0^3}{\tau^3}\frac{{\rm Arctan}\left( \sqrt{\tfrac{\tau_0^2-\tau^2}{\tau^2}} \right)}{\left( \tfrac{\tau_0^2-\tau^2}{\tau^2} \right)^{\tfrac{3}{2}}}.
\end{equation}
In Figure~\ref{Fig_comp} there is a comparison between the integrand appearing in ~(\ref{quantum}), for different $\varepsilon$, and the one appearing in~(\ref{classical}). For relatively small $\varepsilon\le 0.5$ they are almost the same. For larger action scales, larger and larger fluctuations appear. The oscillatory quantum integrals can be computed numerically. Surprisingly enough, the integrated quantum corrections are relatively small (of the order of $1\%$ relative errors or less), up to the values of $\varepsilon\simeq 1$. This value of the relevant action scale of the system $\varepsilon= T_0\tau_0/\hbar$, can correspond, for instance, to a temperature $T_0\simeq 300$ MeV at $\tau_0\simeq 0.67$ fm$/c$. For larger $\varepsilon$ the quantum corrections to the classical values of ${\cal P}_L$ become quickly more important. For $\varepsilon\simeq 2$ (twice the ``starting temperature'' or half the ``starting time'') the $\Delta{\cal P}_L/{\cal P}_L^{\rm class.}$ reaches values of $20\%$. If $\varepsilon\ge3$ one has a very sizable correction $\Delta{\cal P}_L/{\cal P}_L^{\rm class.}\ge50\%$. The numerical computations, if hard and time consuming, can be repeated for different values of $\tau$, and different quantities (the proper energy density ${\cal E}$, the transverse pressure ${\cal P}_T$), with similar outcomes. The quantum corrections tend to be very small for $\varepsilon\ll 1$ and become more and more significant (or even dominant) for $\varepsilon >1$. The corrections tend to be smaller for $\tau \gg \tau_0$ and larger for $\tau \ll \tau_0$. These results suggest that kinetic theory can be a good approximation in the low density limit (almost free-streaming) for the final states~\cite{Borghini:2018xum,Bachmann:2022cls}. While there can be significant quantum corrections to the classical free-streaming, if used in the very initial stages.

\section{Conclusions}
\label{sec:conclusions}

To summarize, in this work it has been studied the free scalar field, as a quantum extension of a non-interacting perfect gas. The evolution equations of the Wigner distributions (the quantum analogue of the distribution function) have been solved exactly for a system symmetric under translations and rotations in the transverse plane. Extending the the results of Refs.~\cite{Akkelin:2018hpu,Akkelin:2020cfs,Rindori:2021quq} to the case of a full $(1+1)$-dimensional expansion. The requirement of local equilibrium for the initial conditions have been relaxed too.

The general solutions of the Wigner distribution are remarkably different from the classical results of kinetic theory. The Wigner distribution is off-shell, and  it shows fast oscillations between positive and negative values. It should be emphasized that such oscillations are always present for any non-trivial, dynamical state in the $(1+1)$-dimensional expansion. It is possible to compute the proper classical limit (on-shell, positive) for a large class of the solutions. The quantum corrections to the classical stress-energy tensor can be relevant (or even dominant) at the early stages of heavy-ion collisions. They are, on the other hand, very small after the freeze-out, as it could have been expected. 

These effects can be phenomenologically relevant whenever a quantum system undergoes a very fast expansion, like in heavy ion collisions. In particular they can be important if the relativistic free streaming is used to connect the initial conditions to  the hydrodyanamic phase.

\section*{Acknowledgements} I would like to thank F. Becattini for the insightful comments. This work has been funded by the Polish National Science Centre grant SONATA 2020/39/D/ST2/02054.

\appendix

\section{Exact evolution of the Wigner distribution for free scalar fields}\label{app:derivation}
Both of the equations~(\ref{Wigner_evolution}) can be found in the literature, for instance in Ref.~\cite{degroot}. However, they are not often presented together. They are usually written with the explicit reference to the sources in the interacting case, which have to be set to zero for free fields. On the other hand, it is rather quick to derive the exact evolution in the compact form of Eq.~(\ref{Wigner_evolution_compact}).

In this appendix it is presented the, probably, most immediate approach. It follows the prescription in Ref.~\cite{Biro:2019jsp} for the Wigner transform of the green functions (section 2.2 in the book). Differently from the case of the Green's functions, applying the Klein-Gordon operator $\nabla +m^2$ to the free fields gives just a number, $0$, instead of a distribution. For the rest, the algebra of the Wigner transform is the same.

Equation~(\ref{Wigner_charged}) can be rewritten in the general way

\begin{equation}\label{Wigner_gen}
    W(x,k)= (1+\kappa)\int \frac{d^4v}{(2\pi)^4} \;  e^{-ik\cdot v} \; {\rm tr}\left( \hat{\rho} : \widehat{\phi}^\dagger(x+\tfrac{v}{2}) \widehat{\phi}(x-\tfrac{v}{2}): \right),
\end{equation}
for both charged and uncharged fields. For charged fields $\kappa=1$. For the uncharged ones $\kappa=0$ and of course the fields are Hermitian $\widehat{\phi}^\dagger = \widehat{\phi}$. The ${\widehat \phi}^\dagger$ in Eq.~(\ref{Wigner_gen}) is unnecessary in this case, but not wrong.

Without loss of generality, one can call $x=(x_1+x_2)/2$. Multiplying Eq.~(\ref{Wigner_gen}) by $e^{ik\cdot(x_1-x_2)}$ and integrating in $d^4k$ one gets, for any distance $x_1-x_2$

\begin{equation}\label{x_1x_2}
    \int d^4k \;  e^{ik\cdot(x_1-x_2)} \; W(\tfrac{x_1+x_2}{2},k) = (1+\kappa)\; {\rm tr}\left( \hat{\rho} : \widehat{\phi}^\dagger(x_1) \widehat{\phi}(x_2): \right).
\end{equation}
Applying the Klein-Gordon operator with respect, eg, to $x_1$ to the right hand side of Eq.~(\ref{x_1x_2}), one has

\begin{equation}
  (1+\kappa)\left( \vphantom{\frac{}{}}\Box_{(x_1)} - m^2 \right) {\rm tr}\left( \hat{\rho} : \widehat{\phi}^\dagger(x_1) \widehat{\phi}(x_2): \right) =  (1+\kappa){\rm tr}\left( \hat{\rho} : \left( \vphantom{\frac{}{}}\Box_{(x_1)} - m^2 \right)\widehat{\phi}^\dagger(x_1) \widehat{\phi}(x_2): \right) =0,
\end{equation}
therefore

\begin{equation}\label{fourier}
    0=\left( \vphantom{\frac{}{}}\Box_{(x_1)} - m^2 \right)\int d^4k \;  e^{ik\cdot(x_1-x_2)} \; W(x=\tfrac{x_1+x_2}{2},k) = \int d^4k \;  e^{ik\cdot(x_1-x_2)} \left[ \frac{1}{4}\Box_{(x)} -k^2 +m^2 +ik\cdot \partial \right] W(x,k).
\end{equation}
To finish the proof one has to make, essentially, a counter-Fourier transform. Namely, multiplying the function $e^{-ik^\prime\cdot(x_1-x_2)}/(2\pi)^4$ and integrating over the $x_1-x_2$. It is immediate to recognize the integral representation of a Dirac delta in 

\begin{equation}
    \int \frac{d^4(x_1-x_2)}{(2\pi)^4} \; e^{-i(k^\prime-k\cdot(x_1-x_2)}= \delta^4(k^\prime -k),
\end{equation}
and finally one has from Eq.~(\ref{fourier})

\begin{equation}
    0=\int d^4k \; \delta^4(k-k^\prime) \left[ \frac{1}{4}\Box -k^2 +m^2 +ik\cdot \partial \right] W(x,k) = \left[ \frac{1}{4}\Box -(k^\prime)^2 +m^2 +ik^\prime\cdot \partial \right] W(x,k^\prime).
\end{equation}
%

\section{Additional constraints for the Wigner distribution}
\label{app:constraints}

Besides the overdetermined set of evolution equations~(\ref{Wigner_evolution}), there are other sources of constraints, for instance the (normal ordered) Hamiltonian has only positive eigenvalues and $0$ for the Minkowski vacuum. The total energy must be positive if there is a system at all. Other, less familiar, constraints can be proved.

The definition itself~(\ref{Wigner_charged}), and the relation between the density matrix $\hat \rho$ and the microscopic states can be used to obtain additional information.

Using the Peskin-Schroeder`s~\cite{Peskin:1995ev} notation and normalization

\begin{equation}
    {\widehat \phi}(x) = \int\frac{d^3 p}{(2\pi)^3\sqrt{2E_{\bf p}}}\left[ \vphantom{\frac{}{}} a_{\bf p} \, e^{-ip\cdot x} \;  + \; b^\dagger_{\bf p}\,  e^{ip\cdot x}  \right],
\end{equation}
one can write directly the Wigner distribution~(\ref{Wigner_charged}) in terms of the expectation value of the creation and annihilation operators

\begin{equation}
    \begin{split}
        W(x,k) &= 2\int\frac{d^4v}{(2\pi)^4}\; e^{-ik\cdot v}\int \frac{d^3p \; d^3q}{2(2\pi)^6\sqrt{E_{\bf p}E_{\bf q}}} \left\{ \langle a^\dagger_{\bf p}a_{\bf q}\rangle \; e^{i(p-q)\cdot x}\;e^{iv\cdot\left(\tfrac{p+q}{2}\right)} \; + \; \langle a^\dagger_{\bf p}b^\dagger_{\bf q}\rangle \; e^{i(p+q)\cdot x}\;e^{iv\cdot\left(\tfrac{p-q}{2}\right)} \right. \\
        & \qquad \left. + \langle b_{\bf p}a_{\bf q}\rangle \; e^{-i(p+q)\cdot x}\;e^{-iv\cdot\left(\tfrac{p-q}{2}\right)} \; + \; \langle b^\dagger_{\bf p}b_{\bf q}\rangle \; e^{i(p-q)\cdot x}\;e^{-iv\cdot\left(\tfrac{p+q}{2}\right)} \right\},
    \end{split}
\end{equation}
in which it has been used the shorthand notation ${\rm tr}(\hat \rho \cdots)= \langle \cdots \rangle$. The $v$ integration yields exclusively to the Dirac deltas $\delta^4(k\pm\frac{p\pm q}{2})$, therefore

\begin{equation}\label{explicit_Wigner}
    \begin{split}
        W(x,k) &= \int \frac{d^3p \; d^3q}{(2\pi)^6\sqrt{E_{\bf p}E_{\bf q}}} \left\{ \vphantom{e^{-iv\cdot\left(\tfrac{p+q}{2}\right)}}\delta^4\left( k-\tfrac{p+q}{2} \right) \;\langle a^\dagger_{\bf p}a_{\bf q}\rangle \; e^{i(p-q)\cdot x}\; + \; \delta^4\left( k+\tfrac{p+q}{2} \right) \; \langle b^\dagger_{\bf p}b_{\bf q}\rangle \; e^{i(p-q)\cdot x} \right. \\
        & \qquad \left. \vphantom{e^{-iv\cdot\left(\tfrac{p+q}{2}\right)}} + \delta^4\left( k+\tfrac{p-q}{2} \right) \;\langle b_{\bf p}a_{\bf q}\rangle \; e^{-i(p+q)\cdot x} \; + \;\delta^4\left( k-\tfrac{p-q}{2} \right) \;\langle a^\dagger_{\bf p}b^\dagger_{\bf q}\rangle \; e^{i(p+q)\cdot x} \right\}.
    \end{split}
\end{equation}
The first two  terms contribute only to the time-like part of $W$, respectively, the particle and antiparticle contribution. More precisely, only to the $k^2\ge m^2$ part. Because of the Dirac deltas, and because of
\begin{equation}
    \begin{split}
        &\left(\frac{p+q}{2}  \right)^2 =\frac{1}{2}\left( \vphantom{\frac{}{}} m^2 +p\cdot q \right).
    \end{split}
\end{equation}
Since both $p$ and $q$ are on-shell, positive energy vectors, with the same mass

\begin{equation}\label{momentum_algebra}
    \begin{split}
        &\left(\frac{p+q}{2}  \right)^2<m^2, \; \Leftrightarrow \; m^2 +p\cdot q <2m^2, \; \Leftrightarrow \; E_{\bf p}E_{\bf q}< m^2 + ({\bf p}\cdot {\bf q}) \; \Rightarrow \; \\
        & \quad\; \Rightarrow \;m^4 + m^2\left( \vphantom{\frac{}{}} |{\bf p}|^2 + |{\bf q}|^2\right) + |{\bf p}|^2|{\bf q}|^2 < m^4 +2 m^2({\bf p}\cdot{\bf q})+ ({\bf p}\cdot{\bf q})^2  \; \Leftrightarrow \; \\
        & \qquad \; \Leftrightarrow \; m^2 \left| \vphantom{\frac{}{}} {\bf p} -{\bf q}\right|^2 < ({\bf p}\cdot{\bf q})^2 - |{\bf p}|^2|{\bf q}|^2 \le 0,
    \end{split}
\end{equation}
which is impossible. Using very similar argument one can prove that the last two terms in~(\ref{explicit_Wigner}) contribute only to the space-like part of $W(x,k)$, since $(\frac{p-q}{2})^2<0$. This has the important implication that not all microscopic states provide a Wigner distribution with a space-like part. The density matrix, pure or mixed, reads

\begin{equation}
    \hat \rho = \sum_i {\sf P}_i |\psi_i\rangle\langle\psi_i|,
\end{equation}
with ${\sf P}_i$ being the (classical $0\le{\sf P}_i\le1$) probability to be in a physical state $|\psi_i\rangle$ (that is $\langle\psi_i||\psi_i\rangle=1$). The extension to an infinite number of possible physical states is immediate. The discrete sum over the classical probability becomes an integral with a probability density.

Only if some of the states $|\psi_i\rangle$ are in a quantum superposition between states differing exactly by one particle-antiparticle pair in particle content the $\langle a^\dagger_{\bf p}b^\dagger_{\bf q}\rangle$ and $\langle a_{\bf p}b_{\bf q}\rangle$ can be different from $0$. In particular, if the states are eigenstates of either the particle number or the antiparticle number operator, these expectation values are vanishing.

Another interesting consequence of~(\ref{explicit_Wigner}) is the fact that the on-shell part of the Wigner distribution is space-time independent, and hence the strictly on-shell Wigner distributions are trivial. Indeed, for any wavevector $k^\mu$, the first four-dimensional Dirac delta in~(\ref{explicit_Wigner}) reads

\begin{equation}
    \delta^4\left( k - \tfrac{p+q}{2} \right)=\delta^3\left({\bf k} -\tfrac{{\bf p}+{\bf q}}{2} \right)\delta\left( k^0 -\tfrac{E_{\bf p}+E_{\bf q}}{2}\right),
\end{equation}
in particular for an on-shell wavevector $k^2=m^2$ ( with positive frequency)

\begin{equation}\label{p=q}
    \begin{split}
    \delta^4\left( k - \tfrac{p+q}{2} \right)&=\delta^3\left({\bf k} -\tfrac{{\bf p}+{\bf q}}{2} \right)\delta\left( \sqrt{m^2 + |{\bf k}|^2} -\tfrac{E_{\bf p}+E_{\bf q}}{2}\right) \\
    &=\delta^3\left({\bf k} -\tfrac{{\bf p}+{\bf q}}{2} \right)\delta\left( \sqrt{m^2 + |\tfrac{{\bf p}+{\bf q}}{2}|^2} -\tfrac{E_{\bf p}+E_{\bf q}}{2}\right).
    \end{split}
\end{equation}
With arguments very similar to the ones used in~(\ref{momentum_algebra}) it is possible, if tedious, to prove that the first delta in the right hand side of the~(\ref{p=q}) is compatible with the second exclusively in the case ${\bf p}={\bf q}$. Looking back at the explicit formula~(\ref{explicit_Wigner}), if ${\bf p}={\bf q}$ the phase $e^{i(p-q)\cdot x}$ that multiply the delta is set to one, removing the space-time dependence,

The analogous formulas for the other Dirac deltas can be obtained in the same way. The generalization to the uncharged scalar field is immediate. The only differences being a division by $2$, and $b_{\bf p}=a_{\bf p}$.


\bibliography{1+1_exact_scalar}

\begin{thebibliography}{50}%
\makeatletter
\providecommand \@ifxundefined [1]{%
 \@ifx{#1\undefined}
}%
\providecommand \@ifnum [1]{%
 \ifnum #1\expandafter \@firstoftwo
 \else \expandafter \@secondoftwo
 \fi
}%
\providecommand \@ifx [1]{%
 \ifx #1\expandafter \@firstoftwo
 \else \expandafter \@secondoftwo
 \fi
}%
\providecommand \natexlab [1]{#1}%
\providecommand \enquote  [1]{``#1''}%
\providecommand \bibnamefont  [1]{#1}%
\providecommand \bibfnamefont [1]{#1}%
\providecommand \citenamefont [1]{#1}%
\providecommand \href@noop [0]{\@secondoftwo}%
\providecommand \href [0]{\begingroup \@sanitize@url \@href}%
\providecommand \@href[1]{\@@startlink{#1}\@@href}%
\providecommand \@@href[1]{\endgroup#1\@@endlink}%
\providecommand \@sanitize@url [0]{\catcode `\\12\catcode `\$12\catcode
  `\&12\catcode `\#12\catcode `\^12\catcode `\_12\catcode `\%12\relax}%
\providecommand \@@startlink[1]{}%
\providecommand \@@endlink[0]{}%
\providecommand \url  [0]{\begingroup\@sanitize@url \@url }%
\providecommand \@url [1]{\endgroup\@href {#1}{\urlprefix }}%
\providecommand \urlprefix  [0]{URL }%
\providecommand \Eprint [0]{\href }%
\providecommand \doibase [0]{https://doi.org/}%
\providecommand \selectlanguage [0]{\@gobble}%
\providecommand \bibinfo  [0]{\@secondoftwo}%
\providecommand \bibfield  [0]{\@secondoftwo}%
\providecommand \translation [1]{[#1]}%
\providecommand \BibitemOpen [0]{}%
\providecommand \bibitemStop [0]{}%
\providecommand \bibitemNoStop [0]{.\EOS\space}%
\providecommand \EOS [0]{\spacefactor3000\relax}%
\providecommand \BibitemShut  [1]{\csname bibitem#1\endcsname}%
\let\auto@bib@innerbib\@empty
\bibitem [{\citenamefont {Kurkela}\ \emph {et~al.}(2019)\citenamefont
  {Kurkela}, \citenamefont {Mazeliauskas}, \citenamefont {Paquet},
  \citenamefont {Schlichting},\ and\ \citenamefont {Teaney}}]{Kurkela:2018wud}%
  \BibitemOpen
  \bibfield  {author} {\bibinfo {author} {\bibfnamefont {A.}~\bibnamefont
  {Kurkela}}, \bibinfo {author} {\bibfnamefont {A.}~\bibnamefont
  {Mazeliauskas}}, \bibinfo {author} {\bibfnamefont {J.-F.}\ \bibnamefont
  {Paquet}}, \bibinfo {author} {\bibfnamefont {S.}~\bibnamefont
  {Schlichting}},\ and\ \bibinfo {author} {\bibfnamefont {D.}~\bibnamefont
  {Teaney}},\ }\bibfield  {title} {\bibinfo {title} {{Matching the
  Nonequilibrium Initial Stage of Heavy Ion Collisions to Hydrodynamics with
  QCD Kinetic Theory}},\ }\href
  {https://doi.org/10.1103/PhysRevLett.122.122302} {\bibfield  {journal}
  {\bibinfo  {journal} {Phys. Rev. Lett.}\ }\textbf {\bibinfo {volume} {122}},\
  \bibinfo {pages} {122302} (\bibinfo {year} {2019})},\ \Eprint
  {https://arxiv.org/abs/1805.01604} {arXiv:1805.01604 [hep-ph]} \BibitemShut
  {NoStop}%
\bibitem [{\citenamefont {Schlichting}\ and\ \citenamefont
  {Teaney}(2019)}]{Schlichting:2019abc}%
  \BibitemOpen
  \bibfield  {author} {\bibinfo {author} {\bibfnamefont {S.}~\bibnamefont
  {Schlichting}}\ and\ \bibinfo {author} {\bibfnamefont {D.}~\bibnamefont
  {Teaney}},\ }\bibfield  {title} {\bibinfo {title} {{The First fm/c of
  Heavy-Ion Collisions}},\ }\href
  {https://doi.org/10.1146/annurev-nucl-101918-023825} {\bibfield  {journal}
  {\bibinfo  {journal} {Ann. Rev. Nucl. Part. Sci.}\ }\textbf {\bibinfo
  {volume} {69}},\ \bibinfo {pages} {447} (\bibinfo {year} {2019})},\ \Eprint
  {https://arxiv.org/abs/1908.02113} {arXiv:1908.02113 [nucl-th]} \BibitemShut
  {NoStop}%
\bibitem [{\citenamefont {Florkowski}\ \emph {et~al.}(2018)\citenamefont
  {Florkowski}, \citenamefont {Heller},\ and\ \citenamefont
  {Spaliński}}]{Florkowski_2018}%
  \BibitemOpen
  \bibfield  {author} {\bibinfo {author} {\bibfnamefont {W.}~\bibnamefont
  {Florkowski}}, \bibinfo {author} {\bibfnamefont {M.~P.}\ \bibnamefont
  {Heller}},\ and\ \bibinfo {author} {\bibfnamefont {M.}~\bibnamefont
  {Spaliński}},\ }\bibfield  {title} {\bibinfo {title} {New theories of
  relativistic hydrodynamics in the lhc era},\ }\href
  {https://doi.org/10.1088/1361-6633/aaa091} {\bibfield  {journal} {\bibinfo
  {journal} {Reports on Progress in Physics}\ }\textbf {\bibinfo {volume}
  {81}},\ \bibinfo {pages} {046001} (\bibinfo {year} {2018})}\BibitemShut
  {NoStop}%
\bibitem [{\citenamefont {Reichert}\ \emph {et~al.}(2022)\citenamefont
  {Reichert}, \citenamefont {Elz}, \citenamefont {Song}, \citenamefont {Coci},
  \citenamefont {Winn}, \citenamefont {Bratkovskaya}, \citenamefont {Aichelin},
  \citenamefont {Steinheimer},\ and\ \citenamefont {Bleicher}}]{Reichert_2022}%
  \BibitemOpen
  \bibfield  {author} {\bibinfo {author} {\bibfnamefont {T.}~\bibnamefont
  {Reichert}}, \bibinfo {author} {\bibfnamefont {A.}~\bibnamefont {Elz}},
  \bibinfo {author} {\bibfnamefont {T.}~\bibnamefont {Song}}, \bibinfo {author}
  {\bibfnamefont {G.}~\bibnamefont {Coci}}, \bibinfo {author} {\bibfnamefont
  {M.}~\bibnamefont {Winn}}, \bibinfo {author} {\bibfnamefont {E.}~\bibnamefont
  {Bratkovskaya}}, \bibinfo {author} {\bibfnamefont {J.}~\bibnamefont
  {Aichelin}}, \bibinfo {author} {\bibfnamefont {J.}~\bibnamefont
  {Steinheimer}},\ and\ \bibinfo {author} {\bibfnamefont {M.}~\bibnamefont
  {Bleicher}},\ }\bibfield  {title} {\bibinfo {title} {Comparison of heavy ion
  transport simulations: Ag + ag collisions at elab = 1.58a gev},\ }\href
  {https://doi.org/10.1088/1361-6471/ac5dfe} {\bibfield  {journal} {\bibinfo
  {journal} {Journal of Physics G: Nuclear and Particle Physics}\ }\textbf
  {\bibinfo {volume} {49}},\ \bibinfo {pages} {055108} (\bibinfo {year}
  {2022})}\BibitemShut {NoStop}%
\bibitem [{\citenamefont {Martin}\ \emph {et~al.}(2022)\citenamefont {Martin},
  \citenamefont {Noronha-Hostler}, \citenamefont {Elfner}, \citenamefont
  {Hammelmann},\ and\ \citenamefont {Hirayama}}]{martin2022influence}%
  \BibitemOpen
  \bibfield  {author} {\bibinfo {author} {\bibfnamefont {J.~S.~S.}\
  \bibnamefont {Martin}}, \bibinfo {author} {\bibfnamefont {J.}~\bibnamefont
  {Noronha-Hostler}}, \bibinfo {author} {\bibfnamefont {H.}~\bibnamefont
  {Elfner}}, \bibinfo {author} {\bibfnamefont {J.}~\bibnamefont {Hammelmann}},\
  and\ \bibinfo {author} {\bibfnamefont {R.}~\bibnamefont {Hirayama}},\
  }\href@noop {} {\bibinfo {title} {Influence of heavy resonances in smash}}
  (\bibinfo {year} {2022}),\ \Eprint {https://arxiv.org/abs/2207.09607}
  {arXiv:2207.09607 [hep-ph]} \BibitemShut {NoStop}%
\bibitem [{\citenamefont {Vredevoogd}\ and\ \citenamefont
  {Pratt}(2009)}]{PhysRevC.79.044915}%
  \BibitemOpen
  \bibfield  {author} {\bibinfo {author} {\bibfnamefont {J.}~\bibnamefont
  {Vredevoogd}}\ and\ \bibinfo {author} {\bibfnamefont {S.}~\bibnamefont
  {Pratt}},\ }\bibfield  {title} {\bibinfo {title} {Universal flow in the first
  stage of relativistic heavy ion collisions},\ }\href
  {https://doi.org/10.1103/PhysRevC.79.044915} {\bibfield  {journal} {\bibinfo
  {journal} {Phys. Rev. C}\ }\textbf {\bibinfo {volume} {79}},\ \bibinfo
  {pages} {044915} (\bibinfo {year} {2009})}\BibitemShut {NoStop}%
\bibitem [{\citenamefont {Bernhard}\ \emph {et~al.}(2016)\citenamefont
  {Bernhard}, \citenamefont {Moreland}, \citenamefont {Bass}, \citenamefont
  {Liu},\ and\ \citenamefont {Heinz}}]{PhysRevC.94.024907}%
  \BibitemOpen
  \bibfield  {author} {\bibinfo {author} {\bibfnamefont {J.~E.}\ \bibnamefont
  {Bernhard}}, \bibinfo {author} {\bibfnamefont {J.~S.}\ \bibnamefont
  {Moreland}}, \bibinfo {author} {\bibfnamefont {S.~A.}\ \bibnamefont {Bass}},
  \bibinfo {author} {\bibfnamefont {J.}~\bibnamefont {Liu}},\ and\ \bibinfo
  {author} {\bibfnamefont {U.}~\bibnamefont {Heinz}},\ }\bibfield  {title}
  {\bibinfo {title} {Applying bayesian parameter estimation to relativistic
  heavy-ion collisions: Simultaneous characterization of the initial state and
  quark-gluon plasma medium},\ }\href
  {https://doi.org/10.1103/PhysRevC.94.024907} {\bibfield  {journal} {\bibinfo
  {journal} {Phys. Rev. C}\ }\textbf {\bibinfo {volume} {94}},\ \bibinfo
  {pages} {024907} (\bibinfo {year} {2016})}\BibitemShut {NoStop}%
\bibitem [{\citenamefont {Nijs}\ \emph {et~al.}(2021)\citenamefont {Nijs},
  \citenamefont {van~der Schee}, \citenamefont {G\"ursoy},\ and\ \citenamefont
  {Snellings}}]{Nijs:2020roc}%
  \BibitemOpen
  \bibfield  {author} {\bibinfo {author} {\bibfnamefont {G.}~\bibnamefont
  {Nijs}}, \bibinfo {author} {\bibfnamefont {W.}~\bibnamefont {van~der Schee}},
  \bibinfo {author} {\bibfnamefont {U.}~\bibnamefont {G\"ursoy}},\ and\
  \bibinfo {author} {\bibfnamefont {R.}~\bibnamefont {Snellings}},\ }\bibfield
  {title} {\bibinfo {title} {{Bayesian analysis of heavy ion collisions with
  the heavy ion computational framework Trajectum}},\ }\href
  {https://doi.org/10.1103/PhysRevC.103.054909} {\bibfield  {journal} {\bibinfo
   {journal} {Phys. Rev. C}\ }\textbf {\bibinfo {volume} {103}},\ \bibinfo
  {pages} {054909} (\bibinfo {year} {2021})},\ \Eprint
  {https://arxiv.org/abs/2010.15134} {arXiv:2010.15134 [nucl-th]} \BibitemShut
  {NoStop}%
\bibitem [{\citenamefont {Parkkila}\ \emph {et~al.}(2022)\citenamefont
  {Parkkila}, \citenamefont {Onnerstad}, \citenamefont {Taghavi}, \citenamefont
  {Mordasini}, \citenamefont {Bilandzic}, \citenamefont {Virta},\ and\
  \citenamefont {Kim}}]{PARKKILA2022137485}%
  \BibitemOpen
  \bibfield  {author} {\bibinfo {author} {\bibfnamefont {J.}~\bibnamefont
  {Parkkila}}, \bibinfo {author} {\bibfnamefont {A.}~\bibnamefont {Onnerstad}},
  \bibinfo {author} {\bibfnamefont {S.}~\bibnamefont {Taghavi}}, \bibinfo
  {author} {\bibfnamefont {C.}~\bibnamefont {Mordasini}}, \bibinfo {author}
  {\bibfnamefont {A.}~\bibnamefont {Bilandzic}}, \bibinfo {author}
  {\bibfnamefont {M.}~\bibnamefont {Virta}},\ and\ \bibinfo {author}
  {\bibfnamefont {D.}~\bibnamefont {Kim}},\ }\bibfield  {title} {\bibinfo
  {title} {New constraints for qcd matter from improved bayesian parameter
  estimation in heavy-ion collisions at lhc},\ }\href
  {https://doi.org/https://doi.org/10.1016/j.physletb.2022.137485} {\bibfield
  {journal} {\bibinfo  {journal} {Physics Letters B}\ }\textbf {\bibinfo
  {volume} {835}},\ \bibinfo {pages} {137485} (\bibinfo {year}
  {2022})}\BibitemShut {NoStop}%
\bibitem [{\citenamefont {Groot}(1980)}]{degroot}%
  \BibitemOpen
  \bibfield  {author} {\bibinfo {author} {\bibfnamefont {S.~R.~D.}\
  \bibnamefont {Groot}},\ }\href@noop {} {\emph {\bibinfo {title} {Relativistic
  Kinetic Theory. Principles and Applications}}}\ (\bibinfo  {publisher}
  {Amsterdam, Netherlands: North-holland ( 1980) 417p},\ \bibinfo {year}
  {1980})\BibitemShut {NoStop}%
\bibitem [{\citenamefont {Becattini}\ and\ \citenamefont
  {Tinti}(2011)}]{Becattini:2011ev}%
  \BibitemOpen
  \bibfield  {author} {\bibinfo {author} {\bibfnamefont {F.}~\bibnamefont
  {Becattini}}\ and\ \bibinfo {author} {\bibfnamefont {L.}~\bibnamefont
  {Tinti}},\ }\bibfield  {title} {\bibinfo {title} {{Thermodynamical
  inequivalence of quantum stress-energy and spin tensors}},\ }\href
  {https://doi.org/10.1103/PhysRevD.84.025013} {\bibfield  {journal} {\bibinfo
  {journal} {Phys. Rev. D}\ }\textbf {\bibinfo {volume} {84}},\ \bibinfo
  {pages} {025013} (\bibinfo {year} {2011})},\ \Eprint
  {https://arxiv.org/abs/1101.5251} {arXiv:1101.5251 [hep-th]} \BibitemShut
  {NoStop}%
\bibitem [{\citenamefont {Becattini}\ and\ \citenamefont
  {Grossi}(2015)}]{Becattini:2015nva}%
  \BibitemOpen
  \bibfield  {author} {\bibinfo {author} {\bibfnamefont {F.}~\bibnamefont
  {Becattini}}\ and\ \bibinfo {author} {\bibfnamefont {E.}~\bibnamefont
  {Grossi}},\ }\bibfield  {title} {\bibinfo {title} {{Quantum corrections to
  the stress-energy tensor in thermodynamic equilibrium with acceleration}},\
  }\href {https://doi.org/10.1103/PhysRevD.92.045037} {\bibfield  {journal}
  {\bibinfo  {journal} {Phys. Rev. D}\ }\textbf {\bibinfo {volume} {92}},\
  \bibinfo {pages} {045037} (\bibinfo {year} {2015})},\ \Eprint
  {https://arxiv.org/abs/1505.07760} {arXiv:1505.07760 [gr-qc]} \BibitemShut
  {NoStop}%
\bibitem [{\citenamefont {Becattini}\ \emph
  {et~al.}(2021{\natexlab{a}})\citenamefont {Becattini}, \citenamefont
  {Buzzegoli},\ and\ \citenamefont {Palermo}}]{Becattini:2020qol}%
  \BibitemOpen
  \bibfield  {author} {\bibinfo {author} {\bibfnamefont {F.}~\bibnamefont
  {Becattini}}, \bibinfo {author} {\bibfnamefont {M.}~\bibnamefont
  {Buzzegoli}},\ and\ \bibinfo {author} {\bibfnamefont {A.}~\bibnamefont
  {Palermo}},\ }\bibfield  {title} {\bibinfo {title} {{Exact equilibrium
  distributions in statistical quantum field theory with rotation and
  acceleration: scalar field}},\ }\href
  {https://doi.org/10.1007/JHEP02(2021)101} {\bibfield  {journal} {\bibinfo
  {journal} {JHEP}\ }\textbf {\bibinfo {volume} {02}},\ \bibinfo {pages}
  {101}},\ \Eprint {https://arxiv.org/abs/2007.08249} {arXiv:2007.08249
  [hep-th]} \BibitemShut {NoStop}%
\bibitem [{\citenamefont {Palermo}\ \emph {et~al.}(2021)\citenamefont
  {Palermo}, \citenamefont {Buzzegoli},\ and\ \citenamefont
  {Becattini}}]{Palermo:2021hlf}%
  \BibitemOpen
  \bibfield  {author} {\bibinfo {author} {\bibfnamefont {A.}~\bibnamefont
  {Palermo}}, \bibinfo {author} {\bibfnamefont {M.}~\bibnamefont {Buzzegoli}},\
  and\ \bibinfo {author} {\bibfnamefont {F.}~\bibnamefont {Becattini}},\
  }\bibfield  {title} {\bibinfo {title} {{Exact equilibrium distributions in
  statistical quantum field theory with rotation and acceleration: Dirac
  field}},\ }\href {https://doi.org/10.1007/JHEP10(2021)077} {\bibfield
  {journal} {\bibinfo  {journal} {JHEP}\ }\textbf {\bibinfo {volume} {10}},\
  \bibinfo {pages} {077}},\ \Eprint {https://arxiv.org/abs/2106.08340}
  {arXiv:2106.08340 [hep-th]} \BibitemShut {NoStop}%
\bibitem [{\citenamefont {Ambru{\c{s} }}\ and\ \citenamefont
  {Winstanley}(2021)}]{Ambrus:2019cvr}%
  \BibitemOpen
  \bibfield  {author} {\bibinfo {author} {\bibfnamefont {V.~E.}\ \bibnamefont
  {Ambru{\c{s} }}}\ and\ \bibinfo {author} {\bibfnamefont {E.}~\bibnamefont
  {Winstanley}},\ }\bibfield  {title} {\bibinfo {title} {Exact solutions in
  quantum field theory under rotation},\ }in\ \href
  {https://doi.org/10.1007/978-3-030-71427-7_4} {\emph {\bibinfo {booktitle}
  {Strongly Interacting Matter under Rotation}}}\ (\bibinfo  {publisher}
  {Springer International Publishing},\ \bibinfo {year} {2021})\ pp.\ \bibinfo
  {pages} {95--135}\BibitemShut {NoStop}%
\bibitem [{\citenamefont {Becattini}\ \emph {et~al.}(2013)\citenamefont
  {Becattini}, \citenamefont {Chandra}, \citenamefont {Del~Zanna},\ and\
  \citenamefont {Grossi}}]{Becattini:2013fla}%
  \BibitemOpen
  \bibfield  {author} {\bibinfo {author} {\bibfnamefont {F.}~\bibnamefont
  {Becattini}}, \bibinfo {author} {\bibfnamefont {V.}~\bibnamefont {Chandra}},
  \bibinfo {author} {\bibfnamefont {L.}~\bibnamefont {Del~Zanna}},\ and\
  \bibinfo {author} {\bibfnamefont {E.}~\bibnamefont {Grossi}},\ }\bibfield
  {title} {\bibinfo {title} {{Relativistic distribution function for particles
  with spin at local thermodynamical equilibrium}},\ }\href
  {https://doi.org/10.1016/j.aop.2013.07.004} {\bibfield  {journal} {\bibinfo
  {journal} {Annals Phys.}\ }\textbf {\bibinfo {volume} {338}},\ \bibinfo
  {pages} {32} (\bibinfo {year} {2013})},\ \Eprint
  {https://arxiv.org/abs/1303.3431} {arXiv:1303.3431 [nucl-th]} \BibitemShut
  {NoStop}%
\bibitem [{\citenamefont {Becattini}\ \emph
  {et~al.}(2021{\natexlab{b}})\citenamefont {Becattini}, \citenamefont
  {Buzzegoli},\ and\ \citenamefont {Palermo}}]{Becattini:2021suc}%
  \BibitemOpen
  \bibfield  {author} {\bibinfo {author} {\bibfnamefont {F.}~\bibnamefont
  {Becattini}}, \bibinfo {author} {\bibfnamefont {M.}~\bibnamefont
  {Buzzegoli}},\ and\ \bibinfo {author} {\bibfnamefont {A.}~\bibnamefont
  {Palermo}},\ }\bibfield  {title} {\bibinfo {title} {{Spin-thermal shear
  coupling in a relativistic fluid}},\ }\href
  {https://doi.org/10.1016/j.physletb.2021.136519} {\bibfield  {journal}
  {\bibinfo  {journal} {Phys. Lett. B}\ }\textbf {\bibinfo {volume} {820}},\
  \bibinfo {pages} {136519} (\bibinfo {year} {2021}{\natexlab{b}})},\ \Eprint
  {https://arxiv.org/abs/2103.10917} {arXiv:2103.10917 [nucl-th]} \BibitemShut
  {NoStop}%
\bibitem [{\citenamefont {Becattini}\ \emph
  {et~al.}(2021{\natexlab{c}})\citenamefont {Becattini}, \citenamefont
  {Buzzegoli}, \citenamefont {Inghirami}, \citenamefont {Karpenko},\ and\
  \citenamefont {Palermo}}]{Becattini:2021iol}%
  \BibitemOpen
  \bibfield  {author} {\bibinfo {author} {\bibfnamefont {F.}~\bibnamefont
  {Becattini}}, \bibinfo {author} {\bibfnamefont {M.}~\bibnamefont
  {Buzzegoli}}, \bibinfo {author} {\bibfnamefont {G.}~\bibnamefont
  {Inghirami}}, \bibinfo {author} {\bibfnamefont {I.}~\bibnamefont
  {Karpenko}},\ and\ \bibinfo {author} {\bibfnamefont {A.}~\bibnamefont
  {Palermo}},\ }\bibfield  {title} {\bibinfo {title} {{Local Polarization and
  Isothermal Local Equilibrium in Relativistic Heavy Ion Collisions}},\ }\href
  {https://doi.org/10.1103/PhysRevLett.127.272302} {\bibfield  {journal}
  {\bibinfo  {journal} {Phys. Rev. Lett.}\ }\textbf {\bibinfo {volume} {127}},\
  \bibinfo {pages} {272302} (\bibinfo {year} {2021}{\natexlab{c}})},\ \Eprint
  {https://arxiv.org/abs/2103.14621} {arXiv:2103.14621 [nucl-th]} \BibitemShut
  {NoStop}%
\bibitem [{\citenamefont {Akkelin}(2019)}]{Akkelin:2018hpu}%
  \BibitemOpen
  \bibfield  {author} {\bibinfo {author} {\bibfnamefont {S.~V.}\ \bibnamefont
  {Akkelin}},\ }\bibfield  {title} {\bibinfo {title} {{Quasi equilibrium state
  of expanding quantum fields and two-pion Bose-Einstein correlations in $pp$
  collisions at the LHC}},\ }\href {https://doi.org/10.1140/epja/i2019-12755-9}
  {\bibfield  {journal} {\bibinfo  {journal} {Eur. Phys. J. A}\ }\textbf
  {\bibinfo {volume} {55}},\ \bibinfo {pages} {78} (\bibinfo {year} {2019})},\
  \Eprint {https://arxiv.org/abs/1812.03905} {arXiv:1812.03905 [hep-ph]}
  \BibitemShut {NoStop}%
\bibitem [{\citenamefont {Akkelin}(2021)}]{Akkelin:2020cfs}%
  \BibitemOpen
  \bibfield  {author} {\bibinfo {author} {\bibfnamefont {S.~V.}\ \bibnamefont
  {Akkelin}},\ }\bibfield  {title} {\bibinfo {title} {{Cosmological particle
  creation in the little bang}},\ }\href
  {https://doi.org/10.1103/PhysRevD.103.116014} {\bibfield  {journal} {\bibinfo
   {journal} {Phys. Rev. D}\ }\textbf {\bibinfo {volume} {103}},\ \bibinfo
  {pages} {116014} (\bibinfo {year} {2021})},\ \Eprint
  {https://arxiv.org/abs/2008.13606} {arXiv:2008.13606 [hep-ph]} \BibitemShut
  {NoStop}%
\bibitem [{\citenamefont {Rindori}\ \emph {et~al.}(2022)\citenamefont
  {Rindori}, \citenamefont {Tinti}, \citenamefont {Becattini},\ and\
  \citenamefont {Rischke}}]{Rindori:2021quq}%
  \BibitemOpen
  \bibfield  {author} {\bibinfo {author} {\bibfnamefont {D.}~\bibnamefont
  {Rindori}}, \bibinfo {author} {\bibfnamefont {L.}~\bibnamefont {Tinti}},
  \bibinfo {author} {\bibfnamefont {F.}~\bibnamefont {Becattini}},\ and\
  \bibinfo {author} {\bibfnamefont {D.~H.}\ \bibnamefont {Rischke}},\
  }\bibfield  {title} {\bibinfo {title} {{Relativistic quantum fluid with boost
  invariance}},\ }\href {https://doi.org/10.1103/PhysRevD.105.056003}
  {\bibfield  {journal} {\bibinfo  {journal} {Phys. Rev. D}\ }\textbf {\bibinfo
  {volume} {105}},\ \bibinfo {pages} {056003} (\bibinfo {year} {2022})},\
  \Eprint {https://arxiv.org/abs/2102.09016} {arXiv:2102.09016 [hep-th]}
  \BibitemShut {NoStop}%
\bibitem [{\citenamefont {Zubarev}\ \emph {et~al.}(1979)\citenamefont
  {Zubarev}, \citenamefont {Prozorkevich},\ and\ \citenamefont
  {Smolyanskii}}]{JOUR}%
  \BibitemOpen
  \bibfield  {author} {\bibinfo {author} {\bibfnamefont {D.~N.}\ \bibnamefont
  {Zubarev}}, \bibinfo {author} {\bibfnamefont {A.~V.}\ \bibnamefont
  {Prozorkevich}},\ and\ \bibinfo {author} {\bibfnamefont {S.~A.}\ \bibnamefont
  {Smolyanskii}},\ }\bibfield  {title} {\bibinfo {title} {Derivation of
  nonlinear generalized equations of quantum relativistic hydrodynamics},\
  }\href {https://doi.org/https://doi.org/10.1007/BF01032069} {\bibfield
  {journal} {\bibinfo  {journal} {Theoretical and Mathematical Physics}\
  }\textbf {\bibinfo {volume} {40}},\ \bibinfo {pages} {821} (\bibinfo {year}
  {1979})}\BibitemShut {NoStop}%
\bibitem [{\citenamefont {{van Weert}}(1982)}]{VANWEERT1982133}%
  \BibitemOpen
  \bibfield  {author} {\bibinfo {author} {\bibfnamefont {C.}~\bibnamefont {{van
  Weert}}},\ }\bibfield  {title} {\bibinfo {title} {Maximum entropy principle
  and relativistic hydrodynamics},\ }\href
  {https://doi.org/https://doi.org/10.1016/0003-4916(82)90338-4} {\bibfield
  {journal} {\bibinfo  {journal} {Annals of Physics}\ }\textbf {\bibinfo
  {volume} {140}},\ \bibinfo {pages} {133} (\bibinfo {year}
  {1982})}\BibitemShut {NoStop}%
\bibitem [{\citenamefont {Becattini}\ \emph {et~al.}(2015)\citenamefont
  {Becattini}, \citenamefont {Bucciantini}, \citenamefont {Grossi},\ and\
  \citenamefont {Tinti}}]{Becattini:2014yxa}%
  \BibitemOpen
  \bibfield  {author} {\bibinfo {author} {\bibfnamefont {F.}~\bibnamefont
  {Becattini}}, \bibinfo {author} {\bibfnamefont {L.}~\bibnamefont
  {Bucciantini}}, \bibinfo {author} {\bibfnamefont {E.}~\bibnamefont
  {Grossi}},\ and\ \bibinfo {author} {\bibfnamefont {L.}~\bibnamefont
  {Tinti}},\ }\bibfield  {title} {\bibinfo {title} {{Local thermodynamical
  equilibrium and the beta frame for a quantum relativistic fluid}},\ }\href
  {https://doi.org/10.1140/epjc/s10052-015-3384-y} {\bibfield  {journal}
  {\bibinfo  {journal} {Eur. Phys. J. C}\ }\textbf {\bibinfo {volume} {75}},\
  \bibinfo {pages} {191} (\bibinfo {year} {2015})},\ \Eprint
  {https://arxiv.org/abs/1403.6265} {arXiv:1403.6265 [hep-th]} \BibitemShut
  {NoStop}%
\bibitem [{\citenamefont {Hayata}\ \emph {et~al.}(2015)\citenamefont {Hayata},
  \citenamefont {Hidaka}, \citenamefont {Noumi},\ and\ \citenamefont
  {Hongo}}]{PhysRevD.92.065008}%
  \BibitemOpen
  \bibfield  {author} {\bibinfo {author} {\bibfnamefont {T.}~\bibnamefont
  {Hayata}}, \bibinfo {author} {\bibfnamefont {Y.}~\bibnamefont {Hidaka}},
  \bibinfo {author} {\bibfnamefont {T.}~\bibnamefont {Noumi}},\ and\ \bibinfo
  {author} {\bibfnamefont {M.}~\bibnamefont {Hongo}},\ }\bibfield  {title}
  {\bibinfo {title} {Relativistic hydrodynamics from quantum field theory on
  the basis of the generalized gibbs ensemble method},\ }\href
  {https://doi.org/10.1103/PhysRevD.92.065008} {\bibfield  {journal} {\bibinfo
  {journal} {Phys. Rev. D}\ }\textbf {\bibinfo {volume} {92}},\ \bibinfo
  {pages} {065008} (\bibinfo {year} {2015})}\BibitemShut {NoStop}%
\bibitem [{\citenamefont {Harutyunyan}\ \emph {et~al.}(2022)\citenamefont
  {Harutyunyan}, \citenamefont {Sedrakian},\ and\ \citenamefont
  {Rischke}}]{Harutyunyan:2021rmb}%
  \BibitemOpen
  \bibfield  {author} {\bibinfo {author} {\bibfnamefont {A.}~\bibnamefont
  {Harutyunyan}}, \bibinfo {author} {\bibfnamefont {A.}~\bibnamefont
  {Sedrakian}},\ and\ \bibinfo {author} {\bibfnamefont {D.~H.}\ \bibnamefont
  {Rischke}},\ }\bibfield  {title} {\bibinfo {title} {{Relativistic
  second-order dissipative hydrodynamics from Zubarev\textquoteright{}s
  non-equilibrium statistical operator}},\ }\href
  {https://doi.org/10.1016/j.aop.2022.168755} {\bibfield  {journal} {\bibinfo
  {journal} {Annals Phys.}\ }\textbf {\bibinfo {volume} {438}},\ \bibinfo
  {pages} {168755} (\bibinfo {year} {2022})},\ \Eprint
  {https://arxiv.org/abs/2110.04595} {arXiv:2110.04595 [nucl-th]} \BibitemShut
  {NoStop}%
\bibitem [{\citenamefont {Bjorken}(1983)}]{Bjorken:1982qr}%
  \BibitemOpen
  \bibfield  {author} {\bibinfo {author} {\bibfnamefont {J.~D.}\ \bibnamefont
  {Bjorken}},\ }\bibfield  {title} {\bibinfo {title} {{Highly Relativistic
  Nucleus-Nucleus Collisions: The Central Rapidity Region}},\ }\href
  {https://doi.org/10.1103/PhysRevD.27.140} {\bibfield  {journal} {\bibinfo
  {journal} {Phys. Rev. D}\ }\textbf {\bibinfo {volume} {27}},\ \bibinfo
  {pages} {140} (\bibinfo {year} {1983})}\BibitemShut {NoStop}%
\bibitem [{\citenamefont {Gubser}(2010)}]{Gubser:2010ze}%
  \BibitemOpen
  \bibfield  {author} {\bibinfo {author} {\bibfnamefont {S.~S.}\ \bibnamefont
  {Gubser}},\ }\bibfield  {title} {\bibinfo {title} {{Symmetry constraints on
  generalizations of Bjorken flow}},\ }\href
  {https://doi.org/10.1103/PhysRevD.82.085027} {\bibfield  {journal} {\bibinfo
  {journal} {Phys. Rev. D}\ }\textbf {\bibinfo {volume} {82}},\ \bibinfo
  {pages} {085027} (\bibinfo {year} {2010})},\ \Eprint
  {https://arxiv.org/abs/1006.0006} {arXiv:1006.0006 [hep-th]} \BibitemShut
  {NoStop}%
\bibitem [{\citenamefont {Banerjee}\ \emph {et~al.}(1989)\citenamefont
  {Banerjee}, \citenamefont {Bhalerao},\ and\ \citenamefont
  {Ravishankar}}]{BANERJEE198916}%
  \BibitemOpen
  \bibfield  {author} {\bibinfo {author} {\bibfnamefont {B.}~\bibnamefont
  {Banerjee}}, \bibinfo {author} {\bibfnamefont {R.}~\bibnamefont {Bhalerao}},\
  and\ \bibinfo {author} {\bibfnamefont {V.}~\bibnamefont {Ravishankar}},\
  }\bibfield  {title} {\bibinfo {title} {Equilibration of the quark-gluon
  plasma produced in relativistic heavy ion collisions},\ }\href
  {https://doi.org/https://doi.org/10.1016/0370-2693(89)91041-1} {\bibfield
  {journal} {\bibinfo  {journal} {Physics Letters B}\ }\textbf {\bibinfo
  {volume} {224}},\ \bibinfo {pages} {16} (\bibinfo {year} {1989})}\BibitemShut
  {NoStop}%
\bibitem [{\citenamefont {Denicol}\ \emph {et~al.}(2014)\citenamefont
  {Denicol}, \citenamefont {Heinz}, \citenamefont {Martinez}, \citenamefont
  {Noronha},\ and\ \citenamefont {Strickland}}]{PhysRevLett.113.202301}%
  \BibitemOpen
  \bibfield  {author} {\bibinfo {author} {\bibfnamefont {G.~S.}\ \bibnamefont
  {Denicol}}, \bibinfo {author} {\bibfnamefont {U.}~\bibnamefont {Heinz}},
  \bibinfo {author} {\bibfnamefont {M.}~\bibnamefont {Martinez}}, \bibinfo
  {author} {\bibfnamefont {J.}~\bibnamefont {Noronha}},\ and\ \bibinfo {author}
  {\bibfnamefont {M.}~\bibnamefont {Strickland}},\ }\bibfield  {title}
  {\bibinfo {title} {New exact solution of the relativistic boltzmann equation
  and its hydrodynamic limit},\ }\href
  {https://doi.org/10.1103/PhysRevLett.113.202301} {\bibfield  {journal}
  {\bibinfo  {journal} {Phys. Rev. Lett.}\ }\textbf {\bibinfo {volume} {113}},\
  \bibinfo {pages} {202301} (\bibinfo {year} {2014})}\BibitemShut {NoStop}%
\bibitem [{\citenamefont {Florkowski}\ and\ \citenamefont
  {Madetko}(2014)}]{Florkowski:2014txa}%
  \BibitemOpen
  \bibfield  {author} {\bibinfo {author} {\bibfnamefont {W.}~\bibnamefont
  {Florkowski}}\ and\ \bibinfo {author} {\bibfnamefont {O.}~\bibnamefont
  {Madetko}},\ }\bibfield  {title} {\bibinfo {title} {{Kinetic description of
  mixtures of anisotropic fluids}},\ }\href
  {https://doi.org/10.5506/APhysPolB.45.1103} {\bibfield  {journal} {\bibinfo
  {journal} {Acta Phys. Polon. B}\ }\textbf {\bibinfo {volume} {45}},\ \bibinfo
  {pages} {1103} (\bibinfo {year} {2014})},\ \Eprint
  {https://arxiv.org/abs/1402.2401} {arXiv:1402.2401 [nucl-th]} \BibitemShut
  {NoStop}%
\bibitem [{\citenamefont {Florkowski}\ \emph {et~al.}(2013)\citenamefont
  {Florkowski}, \citenamefont {Ryblewski},\ and\ \citenamefont
  {Strickland}}]{Florkowski:2013lya}%
  \BibitemOpen
  \bibfield  {author} {\bibinfo {author} {\bibfnamefont {W.}~\bibnamefont
  {Florkowski}}, \bibinfo {author} {\bibfnamefont {R.}~\bibnamefont
  {Ryblewski}},\ and\ \bibinfo {author} {\bibfnamefont {M.}~\bibnamefont
  {Strickland}},\ }\bibfield  {title} {\bibinfo {title} {{Testing viscous and
  anisotropic hydrodynamics in an exactly solvable case}},\ }\href
  {https://doi.org/10.1103/PhysRevC.88.024903} {\bibfield  {journal} {\bibinfo
  {journal} {Phys. Rev. C}\ }\textbf {\bibinfo {volume} {88}},\ \bibinfo
  {pages} {024903} (\bibinfo {year} {2013})},\ \Eprint
  {https://arxiv.org/abs/1305.7234} {arXiv:1305.7234 [nucl-th]} \BibitemShut
  {NoStop}%
\bibitem [{\citenamefont {Florkowski}\ \emph {et~al.}(2015)\citenamefont
  {Florkowski}, \citenamefont {Maksymiuk}, \citenamefont {Ryblewski},\ and\
  \citenamefont {Tinti}}]{Florkowski:2015cba}%
  \BibitemOpen
  \bibfield  {author} {\bibinfo {author} {\bibfnamefont {W.}~\bibnamefont
  {Florkowski}}, \bibinfo {author} {\bibfnamefont {E.}~\bibnamefont
  {Maksymiuk}}, \bibinfo {author} {\bibfnamefont {R.}~\bibnamefont
  {Ryblewski}},\ and\ \bibinfo {author} {\bibfnamefont {L.}~\bibnamefont
  {Tinti}},\ }\bibfield  {title} {\bibinfo {title} {{Anisotropic hydrodynamics
  for a mixture of quark and gluon fluids}},\ }\href
  {https://doi.org/10.1103/PhysRevC.92.054912} {\bibfield  {journal} {\bibinfo
  {journal} {Phys. Rev. C}\ }\textbf {\bibinfo {volume} {92}},\ \bibinfo
  {pages} {054912} (\bibinfo {year} {2015})},\ \Eprint
  {https://arxiv.org/abs/1508.04534} {arXiv:1508.04534 [nucl-th]} \BibitemShut
  {NoStop}%
\bibitem [{\citenamefont {Bazow}\ \emph {et~al.}(2015)\citenamefont {Bazow},
  \citenamefont {Heinz},\ and\ \citenamefont {Martinez}}]{Bazow:2015cha}%
  \BibitemOpen
  \bibfield  {author} {\bibinfo {author} {\bibfnamefont {D.}~\bibnamefont
  {Bazow}}, \bibinfo {author} {\bibfnamefont {U.~W.}\ \bibnamefont {Heinz}},\
  and\ \bibinfo {author} {\bibfnamefont {M.}~\bibnamefont {Martinez}},\
  }\bibfield  {title} {\bibinfo {title} {{Nonconformal viscous anisotropic
  hydrodynamics}},\ }\href {https://doi.org/10.1103/PhysRevC.91.064903}
  {\bibfield  {journal} {\bibinfo  {journal} {Phys. Rev. C}\ }\textbf {\bibinfo
  {volume} {91}},\ \bibinfo {pages} {064903} (\bibinfo {year} {2015})},\
  \Eprint {https://arxiv.org/abs/1503.07443} {arXiv:1503.07443 [nucl-th]}
  \BibitemShut {NoStop}%
\bibitem [{\citenamefont {Nopoush}\ \emph {et~al.}(2015)\citenamefont
  {Nopoush}, \citenamefont {Ryblewski},\ and\ \citenamefont
  {Strickland}}]{Nopoush:2014qba}%
  \BibitemOpen
  \bibfield  {author} {\bibinfo {author} {\bibfnamefont {M.}~\bibnamefont
  {Nopoush}}, \bibinfo {author} {\bibfnamefont {R.}~\bibnamefont {Ryblewski}},\
  and\ \bibinfo {author} {\bibfnamefont {M.}~\bibnamefont {Strickland}},\
  }\bibfield  {title} {\bibinfo {title} {{Anisotropic hydrodynamics for
  conformal Gubser flow}},\ }\href {https://doi.org/10.1103/PhysRevD.91.045007}
  {\bibfield  {journal} {\bibinfo  {journal} {Phys. Rev. D}\ }\textbf {\bibinfo
  {volume} {91}},\ \bibinfo {pages} {045007} (\bibinfo {year} {2015})},\
  \Eprint {https://arxiv.org/abs/1410.6790} {arXiv:1410.6790 [nucl-th]}
  \BibitemShut {NoStop}%
\bibitem [{\citenamefont {Tinti}\ \emph {et~al.}(2016)\citenamefont {Tinti},
  \citenamefont {Ryblewski}, \citenamefont {Florkowski},\ and\ \citenamefont
  {Strickland}}]{Tinti:2015xra}%
  \BibitemOpen
  \bibfield  {author} {\bibinfo {author} {\bibfnamefont {L.}~\bibnamefont
  {Tinti}}, \bibinfo {author} {\bibfnamefont {R.}~\bibnamefont {Ryblewski}},
  \bibinfo {author} {\bibfnamefont {W.}~\bibnamefont {Florkowski}},\ and\
  \bibinfo {author} {\bibfnamefont {M.}~\bibnamefont {Strickland}},\ }\bibfield
   {title} {\bibinfo {title} {{Testing different formulations of leading-order
  anisotropic hydrodynamics}},\ }\href
  {https://doi.org/10.1016/j.nuclphysa.2015.11.006} {\bibfield  {journal}
  {\bibinfo  {journal} {Nucl. Phys. A}\ }\textbf {\bibinfo {volume} {946}},\
  \bibinfo {pages} {29} (\bibinfo {year} {2016})},\ \Eprint
  {https://arxiv.org/abs/1505.06456} {arXiv:1505.06456 [hep-ph]} \BibitemShut
  {NoStop}%
\bibitem [{\citenamefont {Moln\'ar}\ \emph {et~al.}(2016)\citenamefont
  {Moln\'ar}, \citenamefont {Niemi},\ and\ \citenamefont
  {Rischke}}]{PhysRevD.94.125003}%
  \BibitemOpen
  \bibfield  {author} {\bibinfo {author} {\bibfnamefont {E.}~\bibnamefont
  {Moln\'ar}}, \bibinfo {author} {\bibfnamefont {H.}~\bibnamefont {Niemi}},\
  and\ \bibinfo {author} {\bibfnamefont {D.~H.}\ \bibnamefont {Rischke}},\
  }\bibfield  {title} {\bibinfo {title} {Closing the equations of motion of
  anisotropic fluid dynamics by a judicious choice of a moment of the boltzmann
  equation},\ }\href {https://doi.org/10.1103/PhysRevD.94.125003} {\bibfield
  {journal} {\bibinfo  {journal} {Phys. Rev. D}\ }\textbf {\bibinfo {volume}
  {94}},\ \bibinfo {pages} {125003} (\bibinfo {year} {2016})}\BibitemShut
  {NoStop}%
\bibitem [{\citenamefont {Tinti}\ \emph {et~al.}(2019)\citenamefont {Tinti},
  \citenamefont {Vujanovic}, \citenamefont {Noronha},\ and\ \citenamefont
  {Heinz}}]{Tinti:2018qfb}%
  \BibitemOpen
  \bibfield  {author} {\bibinfo {author} {\bibfnamefont {L.}~\bibnamefont
  {Tinti}}, \bibinfo {author} {\bibfnamefont {G.}~\bibnamefont {Vujanovic}},
  \bibinfo {author} {\bibfnamefont {J.}~\bibnamefont {Noronha}},\ and\ \bibinfo
  {author} {\bibfnamefont {U.}~\bibnamefont {Heinz}},\ }\bibfield  {title}
  {\bibinfo {title} {{Resummed hydrodynamic expansion for a plasma of particles
  interacting with fields}},\ }\href
  {https://doi.org/10.1103/PhysRevD.99.016009} {\bibfield  {journal} {\bibinfo
  {journal} {Phys. Rev. D}\ }\textbf {\bibinfo {volume} {99}},\ \bibinfo
  {pages} {016009} (\bibinfo {year} {2019})},\ \Eprint
  {https://arxiv.org/abs/1808.06436} {arXiv:1808.06436 [nucl-th]} \BibitemShut
  {NoStop}%
\bibitem [{\citenamefont {Jaiswal}\ \emph {et~al.}(2022)\citenamefont
  {Jaiswal}, \citenamefont {Chattopadhyay}, \citenamefont {Du}, \citenamefont
  {Heinz},\ and\ \citenamefont {Pal}}]{Jaiswal:2021uvv}%
  \BibitemOpen
  \bibfield  {author} {\bibinfo {author} {\bibfnamefont {S.}~\bibnamefont
  {Jaiswal}}, \bibinfo {author} {\bibfnamefont {C.}~\bibnamefont
  {Chattopadhyay}}, \bibinfo {author} {\bibfnamefont {L.}~\bibnamefont {Du}},
  \bibinfo {author} {\bibfnamefont {U.}~\bibnamefont {Heinz}},\ and\ \bibinfo
  {author} {\bibfnamefont {S.}~\bibnamefont {Pal}},\ }\bibfield  {title}
  {\bibinfo {title} {{Nonconformal kinetic theory and hydrodynamics for Bjorken
  flow}},\ }\href {https://doi.org/10.1103/PhysRevC.105.024911} {\bibfield
  {journal} {\bibinfo  {journal} {Phys. Rev. C}\ }\textbf {\bibinfo {volume}
  {105}},\ \bibinfo {pages} {024911} (\bibinfo {year} {2022})},\ \Eprint
  {https://arxiv.org/abs/2107.10248} {arXiv:2107.10248 [hep-ph]} \BibitemShut
  {NoStop}%
\bibitem [{\citenamefont {Peskin}\ and\ \citenamefont
  {Schroeder}(1995)}]{Peskin:1995ev}%
  \BibitemOpen
  \bibfield  {author} {\bibinfo {author} {\bibfnamefont {M.~E.}\ \bibnamefont
  {Peskin}}\ and\ \bibinfo {author} {\bibfnamefont {D.~V.}\ \bibnamefont
  {Schroeder}},\ }\href@noop {} {\emph {\bibinfo {title} {{An Introduction to
  quantum field theory}}}}\ (\bibinfo  {publisher} {Addison-Wesley},\ \bibinfo
  {address} {Reading, USA},\ \bibinfo {year} {1995})\BibitemShut {NoStop}%
\bibitem [{\citenamefont {Evans}(2010)}]{evans10}%
  \BibitemOpen
  \bibfield  {author} {\bibinfo {author} {\bibfnamefont {L.~C.}\ \bibnamefont
  {Evans}},\ }\href@noop {} {\emph {\bibinfo {title} {Partial differential
  equations}}}\ (\bibinfo  {publisher} {American Mathematical Society},\
  \bibinfo {address} {Providence, R.I.},\ \bibinfo {year} {2010})\BibitemShut
  {NoStop}%
\bibitem [{\citenamefont {Gradshteyn}\ and\ \citenamefont
  {Ryzhik}(1943)}]{Gradshteyn:1943cpj}%
  \BibitemOpen
  \bibfield  {author} {\bibinfo {author} {\bibfnamefont {I.~S.}\ \bibnamefont
  {Gradshteyn}}\ and\ \bibinfo {author} {\bibfnamefont {I.~M.}\ \bibnamefont
  {Ryzhik}},\ }\href@noop {} {\emph {\bibinfo {title} {{Table of Integrals,
  Series, and Products}}}}\ (\bibinfo {year} {1943})\BibitemShut {NoStop}%
\bibitem [{\citenamefont {Zhou}\ and\ \citenamefont
  {Siadat}(2017)}]{zhou2017notes}%
  \BibitemOpen
  \bibfield  {author} {\bibinfo {author} {\bibfnamefont {K.}~\bibnamefont
  {Zhou}}\ and\ \bibinfo {author} {\bibfnamefont {V.}~\bibnamefont {Siadat}},\
  }\href@noop {} {\bibinfo {title} {Notes on harmonic analysis part i: The
  fourier transform}} (\bibinfo {year} {2017}),\ \Eprint
  {https://arxiv.org/abs/1709.03377} {arXiv:1709.03377 [math.CA]} \BibitemShut
  {NoStop}%
\bibitem [{\citenamefont {Becattini}(2021)}]{Becattini:2020sww}%
  \BibitemOpen
  \bibfield  {author} {\bibinfo {author} {\bibfnamefont {F.}~\bibnamefont
  {Becattini}},\ }\bibfield  {title} {\bibinfo {title} {{Polarization in
  relativistic fluids: a quantum field theoretical derivation}},\ }\href
  {https://doi.org/10.1007/978-3-030-71427-7_2} {\bibfield  {journal} {\bibinfo
   {journal} {Lect. Notes Phys.}\ }\textbf {\bibinfo {volume} {987}},\ \bibinfo
  {pages} {15} (\bibinfo {year} {2021})},\ \Eprint
  {https://arxiv.org/abs/2004.04050} {arXiv:2004.04050 [hep-th]} \BibitemShut
  {NoStop}%
\bibitem [{\citenamefont {Tinti}\ and\ \citenamefont
  {Florkowski}(2021)}]{Tinti:2020gyh}%
  \BibitemOpen
  \bibfield  {author} {\bibinfo {author} {\bibfnamefont {L.}~\bibnamefont
  {Tinti}}\ and\ \bibinfo {author} {\bibfnamefont {W.}~\bibnamefont
  {Florkowski}},\ }\bibfield  {title} {\bibinfo {title} {Particle polarization,
  spin tensor, and the wigner distribution in relativistic systems},\ }in\
  \href {https://doi.org/10.1007/978-3-030-71427-7_5} {\emph {\bibinfo
  {booktitle} {Strongly Interacting Matter under Rotation}}}\ (\bibinfo
  {publisher} {Springer International Publishing},\ \bibinfo {year} {2021})\
  pp.\ \bibinfo {pages} {137--165}\BibitemShut {NoStop}%
\bibitem [{\citenamefont {Wheeden}\ and\ \citenamefont
  {Zygmund}(1977)}]{wheeden1977measure}%
  \BibitemOpen
  \bibfield  {author} {\bibinfo {author} {\bibfnamefont {R.}~\bibnamefont
  {Wheeden}}\ and\ \bibinfo {author} {\bibfnamefont {A.}~\bibnamefont
  {Zygmund}},\ }\href {https://books.google.pl/books?id=DTgOrgEACAAJ} {\emph
  {\bibinfo {title} {Measure and Integral}}},\ (Pure and applied mathematics)\
  (\bibinfo  {publisher} {Dekker},\ \bibinfo {year} {1977})\BibitemShut
  {NoStop}%
\bibitem [{\citenamefont {Tinti}(2020)}]{Tinti:2020pyb}%
  \BibitemOpen
  \bibfield  {author} {\bibinfo {author} {\bibfnamefont {L.}~\bibnamefont
  {Tinti}},\ }\href@noop {} {\bibinfo {title} {Hydrodynamics from quantum
  fields: a regularized expansion from the wigner distribution}} (\bibinfo
  {year} {2020}),\ \Eprint {https://arxiv.org/abs/2003.09268} {arXiv:2003.09268
  [nucl-th]} \BibitemShut {NoStop}%
\bibitem [{\citenamefont {Borghini}\ \emph {et~al.}(2018)\citenamefont
  {Borghini}, \citenamefont {Feld},\ and\ \citenamefont
  {Kersting}}]{Borghini:2018xum}%
  \BibitemOpen
  \bibfield  {author} {\bibinfo {author} {\bibfnamefont {N.}~\bibnamefont
  {Borghini}}, \bibinfo {author} {\bibfnamefont {S.}~\bibnamefont {Feld}},\
  and\ \bibinfo {author} {\bibfnamefont {N.}~\bibnamefont {Kersting}},\
  }\bibfield  {title} {\bibinfo {title} {{Scaling behavior of anisotropic flow
  harmonics in the far from equilibrium regime}},\ }\href
  {https://doi.org/10.1140/epjc/s10052-018-6313-z} {\bibfield  {journal}
  {\bibinfo  {journal} {Eur. Phys. J. C}\ }\textbf {\bibinfo {volume} {78}},\
  \bibinfo {pages} {832} (\bibinfo {year} {2018})},\ \Eprint
  {https://arxiv.org/abs/1804.05729} {arXiv:1804.05729 [nucl-th]} \BibitemShut
  {NoStop}%
\bibitem [{\citenamefont {Bachmann}\ \emph {et~al.}(2023)\citenamefont
  {Bachmann}, \citenamefont {Borghini}, \citenamefont {Feld},\ and\
  \citenamefont {Roch}}]{Bachmann:2022cls}%
  \BibitemOpen
  \bibfield  {author} {\bibinfo {author} {\bibfnamefont {B.}~\bibnamefont
  {Bachmann}}, \bibinfo {author} {\bibfnamefont {N.}~\bibnamefont {Borghini}},
  \bibinfo {author} {\bibfnamefont {N.}~\bibnamefont {Feld}},\ and\ \bibinfo
  {author} {\bibfnamefont {H.}~\bibnamefont {Roch}},\ }\bibfield  {title}
  {\bibinfo {title} {{On differences between even and odd anisotropic-flow
  harmonics in non-equilibrated systems}},\ }\href
  {https://doi.org/10.1140/epjc/s10052-023-11256-w} {\bibfield  {journal}
  {\bibinfo  {journal} {Eur. Phys. J. C}\ }\textbf {\bibinfo {volume} {83}},\
  \bibinfo {pages} {114} (\bibinfo {year} {2023})},\ \Eprint
  {https://arxiv.org/abs/2203.13306} {arXiv:2203.13306 [nucl-th]} \BibitemShut
  {NoStop}%
\bibitem [{\citenamefont {Bir\'o}\ and\ \citenamefont
  {Jakov\'ac}(2019)}]{Biro:2019jsp}%
  \BibitemOpen
  \bibfield  {author} {\bibinfo {author} {\bibfnamefont {T.~S.}\ \bibnamefont
  {Bir\'o}}\ and\ \bibinfo {author} {\bibfnamefont {A.}~\bibnamefont
  {Jakov\'ac}},\ }\href {https://doi.org/10.1007/978-3-030-11689-7} {\emph
  {\bibinfo {title} {{Emergence of Temperature in Examples and Related
  Nuisances in Field Theory}}}},\ SpringerBriefs in Physics\ (\bibinfo
  {publisher} {Springer},\ \bibinfo {year} {2019})\BibitemShut {NoStop}%
\end{thebibliography}%

\end{document}